\documentclass[fleqn,usenatbib]{mnras}
\usepackage{newtxtext,newtxmath}

\usepackage[T1]{fontenc}
\usepackage{ae,aecompl}

\usepackage{graphicx}	
\usepackage{amsmath}	

\usepackage{natbib}

\usepackage{enumerate}
\usepackage{color}
\usepackage{graphicx}
\usepackage{multicol}
\usepackage{savesym}
\usepackage{color}
\usepackage[safe]{tipa}
\usepackage{csvsimple}
\usepackage{arydshln}
\usepackage{arydshln}
\usepackage{hyphenat}

\newcommand{\arcsecond}{$^{\prime\prime}$}

\title[Multiband study of ULXs in NGC\,7424]
  {A multiband look at ultraluminous X-ray sources in NGC\,7424}

\author[R. Soria et al.]
{Roberto Soria$^{1,2,3}$ \thanks{Email: roberto.soria@inaf.it (RS)}, 
Siying Cheng$^1$,
Manfred W.~Pakull$^{4}$,
Christian~Motch$^{4}$,
\newauthor
and Thomas D.~Russell$^{5}$\\
\\
$^{1}$College of Astronomy and Space Sciences, University of the Chinese Academy of Sciences, Beijing 100049, China\\
$^{2}$ INAF-Osservatorio Astrofisico di Torino, Strada Osservatorio 20, I-10025 Pino Torinese, Italy\\
$^{3}$Sydney Institute for Astronomy, School of Physics A28, The University of Sydney, Sydney, NSW 2006, Australia\\
$^{4}$ Universit\'e de Strasbourg, CNRS, Observatoire astronomique, CNRS, UMR 7550,F-67000, Strasbourg, France\\
$^5$INAF, Istituto di Astrofisica Spaziale e Fisica Cosmica, Palermo, Italy\\
}

\date{Accepted 14 February 2024 --- Received 12 February 2024 --- in original form 2 January 2024}

\pubyear{2024}

\begin{document}
\label{firstpage}
\pagerange{\pageref{firstpage}--\pageref{lastpage}}
\maketitle

\begin{abstract}
We studied the multiband properties of two ultraluminous X-ray sources (2CXO J225728.9$-$410211 = X-1 and 2CXO J225724.7$-$410343 = X-2) and their surroundings, in the spiral galaxy NGC\,7424. Both sources have approached X-ray luminosities $L_{\rm X} \sim 10^{40}$ erg s$^{-1}$ at some epochs. Thanks to a more accurate astrometric solution (based on Australia Telescope Compact Array and {\it Gaia} data), we identified the point-like optical counterpart of X-1, which looks like an isolated B8 supergiant ($M \approx 9 M_\odot$, age $\approx 30$ Myr). Instead, X-2 is in a star-forming region (size of about 100 pc $\times$ 150 pc), near young clusters and ionized gas. Very Large Telescope long-slit spectra show a spatially extended region of He {\footnotesize {II}} $\lambda$4686 emission around the X-ray position, displaced by about 50 pc from the brightest star cluster, which corresponds to the peak of lower-ionization line emission. We interpret the He {\footnotesize {II}} $\lambda$4686 emission as a signature of X-ray photo-ionization from the ULX, while the other optical lines are consistent with UV ionization in an ordinary He {\footnotesize {II}} region. The luminosity of this He$^{++}$ nebula puts it in the same class as other classical photo-ionized ULX nebulae such as those around Holmberg II X-1 and NGC\,5408 X-1. We locate a strong (5.5-GHz luminosity $\nu\,L_{\nu} \approx 10^{35}$ erg s$^{-1}$), steep-spectrum, unresolved radio source at the peak of the low-ionization lines, and discuss alternative physical scenarios for the radio emission. Finally, we use {\it WISE} data to obtain an independent estimate of the reddening of the star-forming clump around X-2.   
\end{abstract}



\section{Introduction}

X-ray and multiband studies of the most luminous off-nuclear sources in nearby galaxies (ultraluminous X-ray sources, ULXs; see reviews by   \citealt{pinto23,king23,kaaret17,feng11}) have proved that such sources are usually powered by accretion onto stellar-mass compact objects. In most cases, they are fed by a high mass donor star ({\it i.e.}, they are high mass X-ray binaries) and can reach X-ray luminosities in excess of $10^{40}$ erg s$^{-1}$, well above their critical Eddington limit (super-critical accretion regime). Investigating super-critical stellar-mass sources in the local universe helps our modelling of accretion and feedback properties in this accretion regime at all scales, including for example in the early phases of supermassive black hole growth \citep[{\it e.g.},][]{king15,tombesi15,volonteri15,parker17}.

One of the distinguishing properties of super-critical accretion is the strong effect such sources have on the surrounding medium, particularly as the most powerful sources tend to be located in gas-rich, young stellar environments. X-ray photo-ionization effects are expected because of their high luminosity and long mean-free-path of their X-ray photons. It was also speculated that ULXs may have played a role in cosmic re-ionization \citep[{\it e.g.},][]{mirabel11,fragos13,madau17,douna18}, at least for the pre-heating of the intergalactic medium and the formation of extended partially ionized zones \citep[{\it e.g.},][]{jeon14,knevitt14,xu14}. Moreover, super-critical sources produce strong radiatively-driven outflows \citep[{\it e.g.},][]{king03,poutanen07,dotan11,kosec18,pinto23b}. Magneto-hydrodynamical simulations \citep[{\it e.g.},][]{ohsuga11,kawashima12,jiang14,ogawa17,narayan17,kitaki21} show that the massive wind from a geometrically thick disk produce a lower-density polar funnel, inside which a collimated jet may also be launched. Theoretical studies show \citep{poutanen07,yoshioka22} that the kinetic power of the super-Eddington outflows can reach $\sim$10--30\% of the radiative luminosity, and possibly even exceed the radiative luminosity for accretion rates in excess of about 40 times the critical limit \citep{kitaki21}. This is supported by observational discoveries of large (diameters of $\sim$100--300 pc) shock-ionized bubbles around several ULXs, inflated by mechanical powers of $\sim$10$^{39}$--10$^{40}$ erg s$^{-1}$ \citep{pakull02,pakull10,cseh12,soria21,gurpide22,zhou23}.
In summary, by studying the gas and stellar environment around a ULX, we constrain the age and activity phase of the super-Eddington source, and its ionizing effect on the surrounding gas.

In this paper, we investigate two ULXs in the grand-design, face-on spiral galaxy NGC\,7424 (Figure 1), with morphological classification SABcd. In the absence of Cepheid distances to this galaxy, we adopt the average between the Hubble Flow distance (10.1 Mpc) and the Tully-Fisher distance (11.5 Mpc), both listed in the NASA/IPAC Extragalactic Database (NED) \footnote{\url{https://ned.ipac.caltech.edu}.}; thus, we take $d = 10.8$ Mpc (distance modulus 30.17 mag, scale of 52 pc arcsec$^{-1}$). The star formation rate is $\approx$0.2--0.3 $M_\odot$ yr$^{-1}$ (\citealt{larsen02,iglesias06}, accounting for the slightly lower distance assumed here). NGC\,7424 was host to the Type-IIb supernova SN 2001ig; some of the observational datasets used for this work were originally collected for a study of that SN. We will report on the evolution of SN 2001ig in a separate paper, and focus here instead on the ULXs and their environments.

\begin{figure}
\includegraphics[width=\columnwidth]{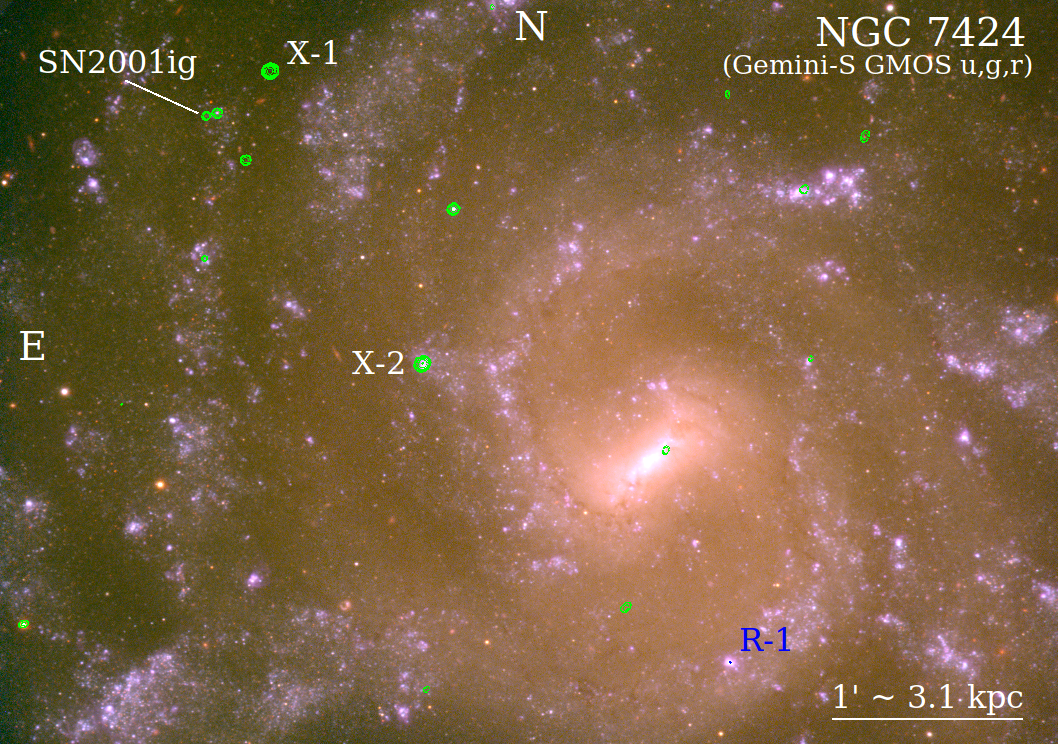}\\[-10pt]
 \caption{Gemini-South GMOS image of NGC\,7424 (red = $r^{\prime}$ filter, green = $g^{\prime}$, blue = $u^{\prime}$), with {\it Chandra}/ACIS-S contours of the brightest X-ray sources (0.3--7 keV band) overplotted in green. We have also labelled the location of a strong, persistent radio source (R-1) without any X-ray counterpart.}
  \label{gemini_full}
\end{figure}

\begin{figure*}
\centering
\includegraphics[height=0.29\textwidth]{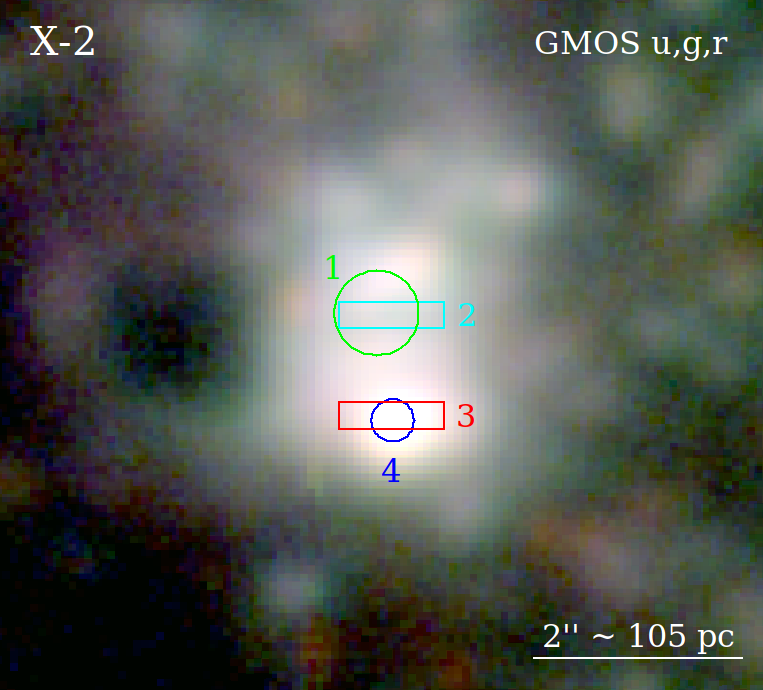}
\includegraphics[height=0.29\textwidth]{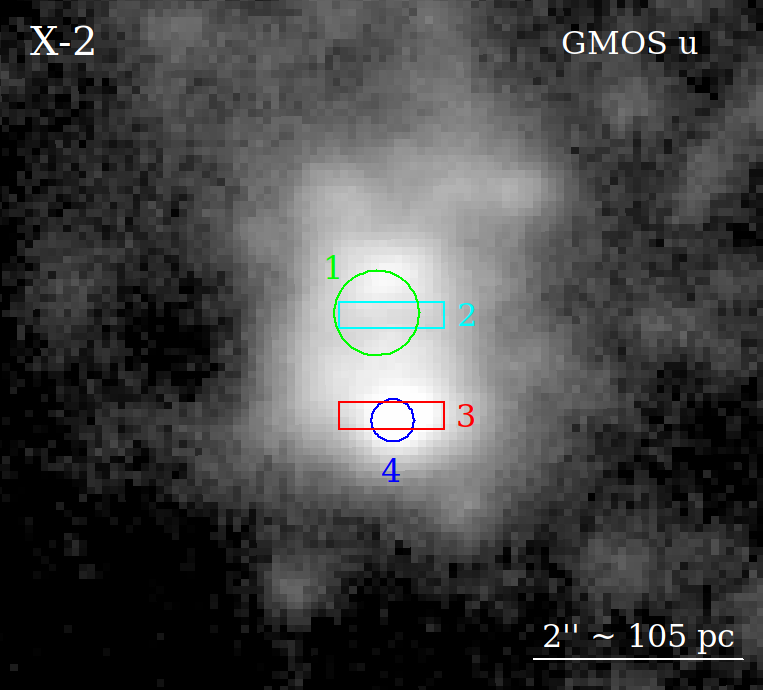}
\includegraphics[height=0.29\textwidth]{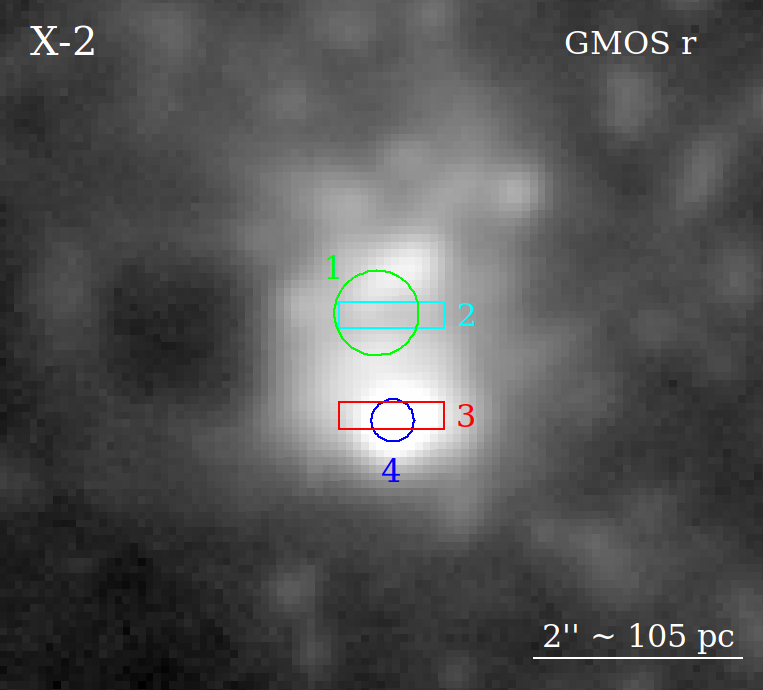}
 \caption{Left panel: zoomed-in true-colour view of the field around X-2, from the Gemini-South GMOS images in the $u^{\prime}$ (blue), $g^{\prime}$ (green) and $r^{\prime}$ (red) filters. Middle panel: GMOS image in the $u^{\prime}$ filter. Right panel: GMOS image in the $r^{\prime}$ filter. In all panels, we have overplotted the location of interesting multiband associations. The green circle (1) is the 90\% confidence limit (radius of 0\farcs4) of the {\it Chandra} position of the ULX X-2. The cyan box (2) is the spatial location along the VLT slit corresponding to the peak in \ion{He}{II} $\lambda$4686 emission; size of the box: 1\arcsecond $\times$ 0\farcs25. The red box (3) corresponds to the peak of the low-ionization lines (\ion{H}{$\alpha$}, \ion{H}{$\beta$}, [\ion{S}{II}]$\lambda\lambda$6716,6731 and several \ion{He}{I} lines).  The blue circle (4) is the 90\% confidence limit (radius of 0\farcs2) of the centroid position of the strong, unresolved radio source detected by the ATCA.}
  \label{gemini_3panels}
\end{figure*}

\section{Observations and data analysis}

\subsection{{\it Chandra}}

NGC\,7424 was observed with the Advanced CCD Imaging Spectrometer (ACIS) camera on board the {\it{Chandra X-ray Observatory}} three times (Table 1): on 2002 May 21 and June 11, and on 2020 December 2 (ObsIDs 3495, 3496, 23572 respectively). The sources of interest for this work were located on the S3 chip. The live time was 23 ks, 24 ks and 5 ks for the three observations.

We retrieved the data from the public archives, and reprocessed them with the Chandra Interactive Analysis of Observations ({\sc{ciao}}) version 4.15 \citep{fruscione06}, with calibration database version 4.9.8. Specifically, we created new level-2 event files with the {\sc ciao} task {\it chandra\_repro}. We used {\it merge\_obs} to create stacked event files and images, and {\it dmcopy} for energy filtering. 
Point-like sources were identified and located with {\it wavdetect}, in the images from each observation.

We used {\it specextract} to create spectra and associated response and ancillary response files (``correctpsf'' parameter set to yes) for the two ULXs. We chose source extraction circles of 2\farcs5 radius, and local background annuli with an area approximately ten times larger.
We used the {\sc ftools}\footnote{\url{http://heasarc.gsfc.nasa.gov/ftools}} package \citep{blackburn95,heasoft14} from NASA's High Energy Astrophysics Science Archive Research Center (HEASARC) for further data analysis. We regrouped the spectra to a minimum number of counts per bin with {\sc ftools}'s {\it grppha} task. In particular, we created spectra with $\geq$1 count per bin, suitable for fitting with the Cash statistics \citep{cash79}, and corresponding spectra grouped to $\geq$15 counts per bin, for $\chi^2$ fitting.

We modelled the X-ray spectra over the 0.3--8 keV band with {\sc{XSPEC}} version 12.12.1 \citep{arnaud96}, with standard models suitable to accreting compact objects. The spectra for some of the observations ({\it e.g.}, X-2 in ObsIDs 3495 and 23573) only have $\sim$100 counts and are mildly undersampled when rebinned for $\chi^2$ fitting; therefore, for consistency, we report the modelling results obtained with the Cash statistics for all spectra, unless explicitly mentioned. Observed fluxes and unabsorbed luminosities were calculated with the {\it cflux} convolution model.

\subsection{Gemini}
NGC\,7424 was imaged with the Gemini Multi-Object Spectrograph (GMOS) on the Gemini South telescope, on 2004 September 14 \citep[][Program ID GS-2004B-Q-6]{Ryder2006}. Conditions were not photometric (thin cirrus clouds) but the seeing was exceptionally good (0\farcs35--0\farcs4). The images were taken in the $u^{\prime},g^{\prime},r^{\prime}$ set of Sloan filters, with a 0\farcs0807/pixel sampling. A series of dithered sub-exposures were taken for each filter, and combined to remove cosmic rays and chip gaps; see \cite{Ryder2006} for details of the observational set-up and data reduction.
The 5\farcm5 $\times$ 5\farcm5 field of view of GMOS covers most of the star-forming disk of NGC\,7424 and is well suited for a search of the optical counterparts to the {\it Chandra} sources (Figure 1).

\subsection{{\it Hubble Space Telescope (HST)}}

For the imaging study of our main target of interest, the optical nebula around X-2, we used an {\it HST} Wide Field Planetary Camera 2 (WFPC2) image in the F606W filter, taken on 1994 July 16 (exposure time 160 s), and a WFPC2 image in the F814W filter taken on 2001 June 1 (exposure time 640 s). To image the star-forming complex around the bright radio source R1 (S.~Cheng et al.,in prep.), in the southern half of the galaxy, we used a different set of WFPC2 images, taken on 2001 July 7, in the F450W filter (460-s exposure) and in the F814W filter (460-s exposure). The latter pair of WFPC2 images do not cover the X-2 field. For a search of the optical counterpart of X-1, we used two Wide Field Camera 3 (WFC3) UVIS images, taken on 2016 April 28 \citep{Ryder2018}: in the F275W (8694-s exposure) and in F336W (2920-s exposure). The field of view of those WFC3/UVIS images includes also the counterpart of SN\,2001ig but not the field around X-2.

We downloaded the drizzled, calibrated data from the Hubble Legacy Archive \footnote{\url{https://hla.stsci.edu}} for the WFPC2 images files, and from the Mikulski Archive for Space Telescopes\footnote{\url{https://mast.stsci.edu/search/ui/{\#}/hst}} for the WFC3 ones.
We used {\sc ds9} imaging tools to determine the centroids of bright point-like sources and improve the astrometric solution (Section 3.1), and, subsequently, for direct comparisons of X-ray, optical and radio positions of our targets of interest.

\begin{figure}
\centering
\includegraphics[width=\columnwidth]{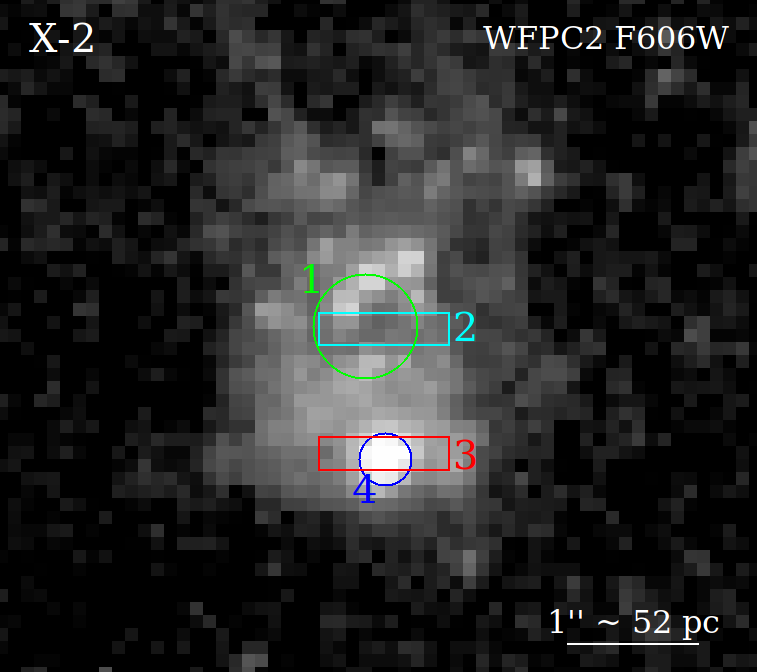}\\
\includegraphics[width=\columnwidth]{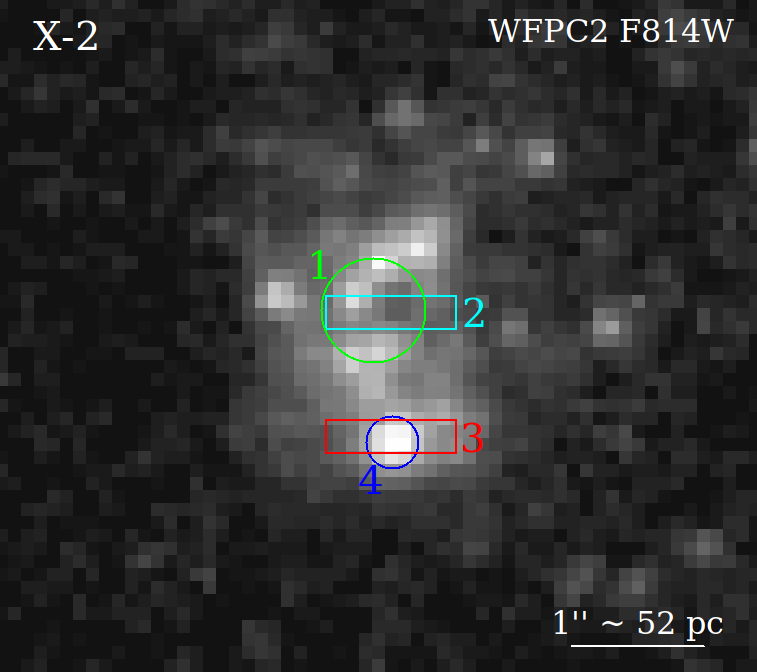}
 \caption{Top panel: {\it HST}/WFPC2 image of the X-2 field, in the F606W filter. Bottom panel: {\it HST}/WFPC2 image in the F814W filter. In both panels, the meaning of the four overplotted regions is the same as in Figure 2.}
  \label{hst_wfpc2}
\end{figure}

\subsection{Very Large Telescope (VLT)}
We observed the X-2 nebula with the FOcal Reducer and low dispersion Spectrograh 2 (FORS2) on the VLT of the European Southern Observatory, on 2011 September 1, from MJD 55805.09 to MJD 55805.11 (corresponding to UT time between about 02:05 and 02:40) (Table 1). Specifically, we took two spectra with the 300V grating (covering 3300 \AA\ to 9500 \AA) and one spectrum with the 1200R grating (covering 5750 \AA\ to 7310 \AA), and an acquisition image before each choice of grating (Table 1). In the spatial direction, the scale was 0\farcs25 per pixel (2 $\times$ 2 pixel binning). The dispersion was $\approx$3.30 \AA\ per pixel for the 300V grating, and 0.75 \AA\ per pixel for the 1200R grating. The instrumental resolution was 9.4 \AA\ full-width-half-maximum (FWHM) for the 300V grism, and 1.8 \AA\ FWHM for the 1200R one. The slit width was 1\farcs0 for the 300V grating (matched to a delivered seeing of 1\farcs1) and 0\farcs7 for the 1200R grating (delivered seeing of 0\farcs9). In both cases, the slit was oriented north to south, passing roughly through the middle of the X-2 nebula.

Spectra were corrected for bias, flat-fielded and calibrated in wavelength and flux using the {\it EsoReflex} FORS package \citep{freudling13} version 5.6.5. Flux calibration was derived from the spectrophotometric standard LTT7379.  We used software from both the Munich Image Data Analysis System ({\sc midas}: \citealt{warmels92}) (in particular, the {\it {integrate/line}} task) and from the Image Reduction and Analysis Facility ({\sc iraf}) Version 2.16. In particular, we used the {\sc iraf} {\it splot} sub-package to determine central wavelengths of the emission lines, their equivalent widths (EWs), FWHM, and fluxes.

\subsection{Australia Telescope Compact Array (ATCA)}

We observed NGC\,7424 with the ATCA on 2021 April 10--13 (Table 1), under project code C3421. The aimpoint was near the position of X-2. The array was in its most extended 6 km configuration (6D)\footnote{\url{https://www.narrabri.atnf.csiro.au/operations/array\_configurations/configurations.html}}. The data were recorded simultaneously at central frequencies of 5.5 and 9.0 GHz, with a bandwidth of 2 GHz at each frequency. The total observing time on source was 21 hr. We used the primary ATCA calibrator PKS B1934$-$638 for flux density calibration, and the nearby source PKS B2310$-$417 for phase calibration. We processed data following standard procedures\footnote{\url{https://casaguides.nrao.edu/index.php/ATCA\_Tutorials}} within the Common Astronomy Software Application ({\sc casa}, version 5.1.2; \citealt{casa22}). To image the data we used the \textsc{casa} task \textit{clean}, using a Briggs robust parameter of 0 \citep{briggs95}, balancing sensitivity and resolution. 
These choices resulted in synthesized beams with FWHM of 2\farcs5 $\times$ 1\farcs4 at 5.5 GHz (position angle $+1^{\circ}$, east of north), and 1\farcs5 $\times$ 0\farcs9 at 9 GHz (position angle $-2^{\circ}$). 
To determine the centroids of our sources of interest, we used the {\sc casa} task {\it imfit}, fitting a 2-D Gaussian with a FWHM fixed to the parameters of the synthesized beam.

\begin{figure}
\centering
\includegraphics[width=\columnwidth]{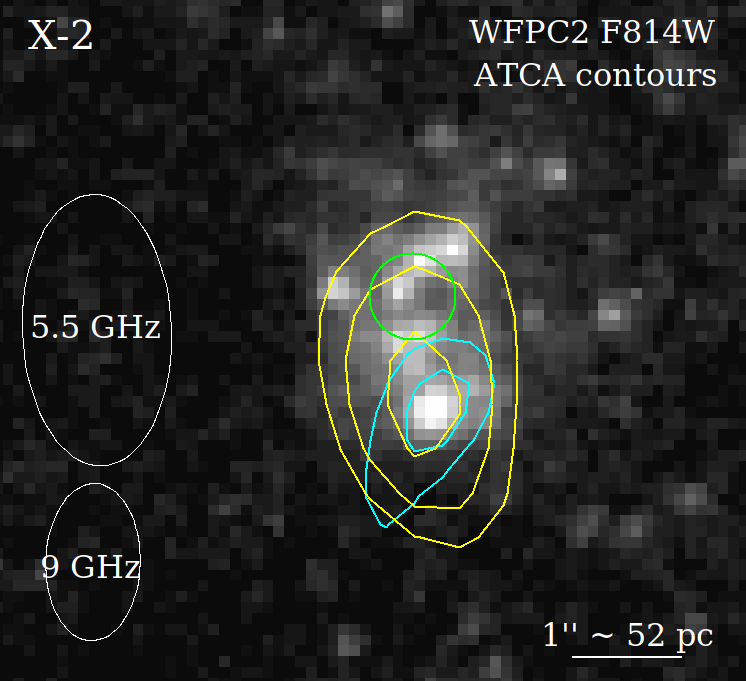}
\caption{ATCA flux density contours overplotted on the {\it HST}/WFPC2 F814W image. Cyan contours are for 9 GHz, yellow contours for 5.5 GHz. Contours are defined as $2^{n/2}$ times the local rms noise level. For the 9 GHz map, the two plotted contours correspond to 40 $\mu$Jy (4$\sigma$) and 57 $\mu$Jy (5.7$\sigma$). For the 5.5 GHz map, the three plotted contours correspond to 60 $\mu$Jy (4$\sigma$), 85 $\mu$Jy (5.7$\sigma$) and 120 $\mu$Jy (8$\sigma$). The {\it Chandra} position of X-2 is overplotted as a green circle. North is up, East to the left.
}

\label{atca_hst}
\end{figure}

\begin{figure}
\vspace{-0.7cm}
\hspace{-0.4cm}
\includegraphics[height=1.25\columnwidth, angle=270]{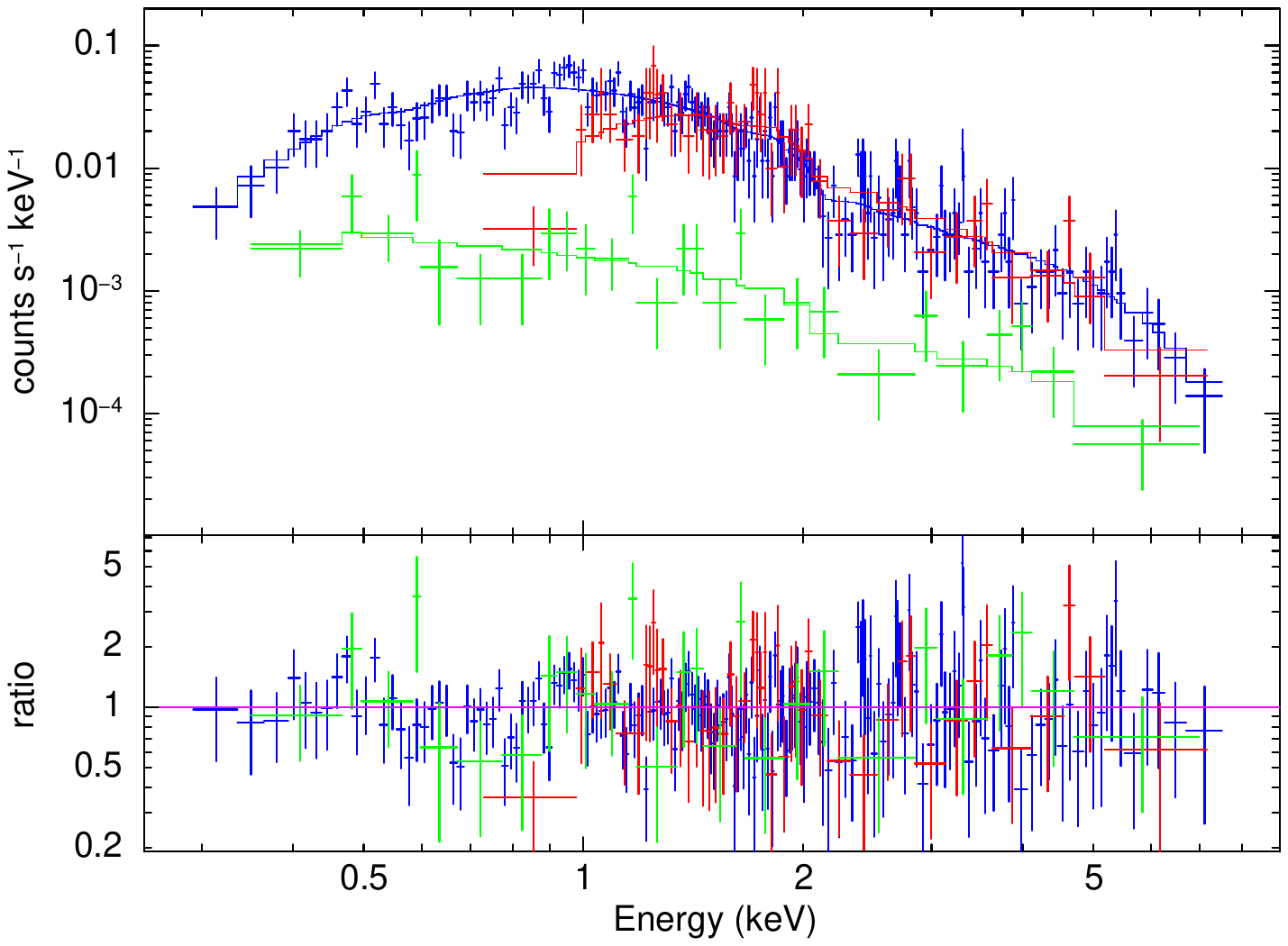}\\[-35pt]

\hspace{-0.4cm}
\includegraphics[height=1.25\columnwidth, angle=270]{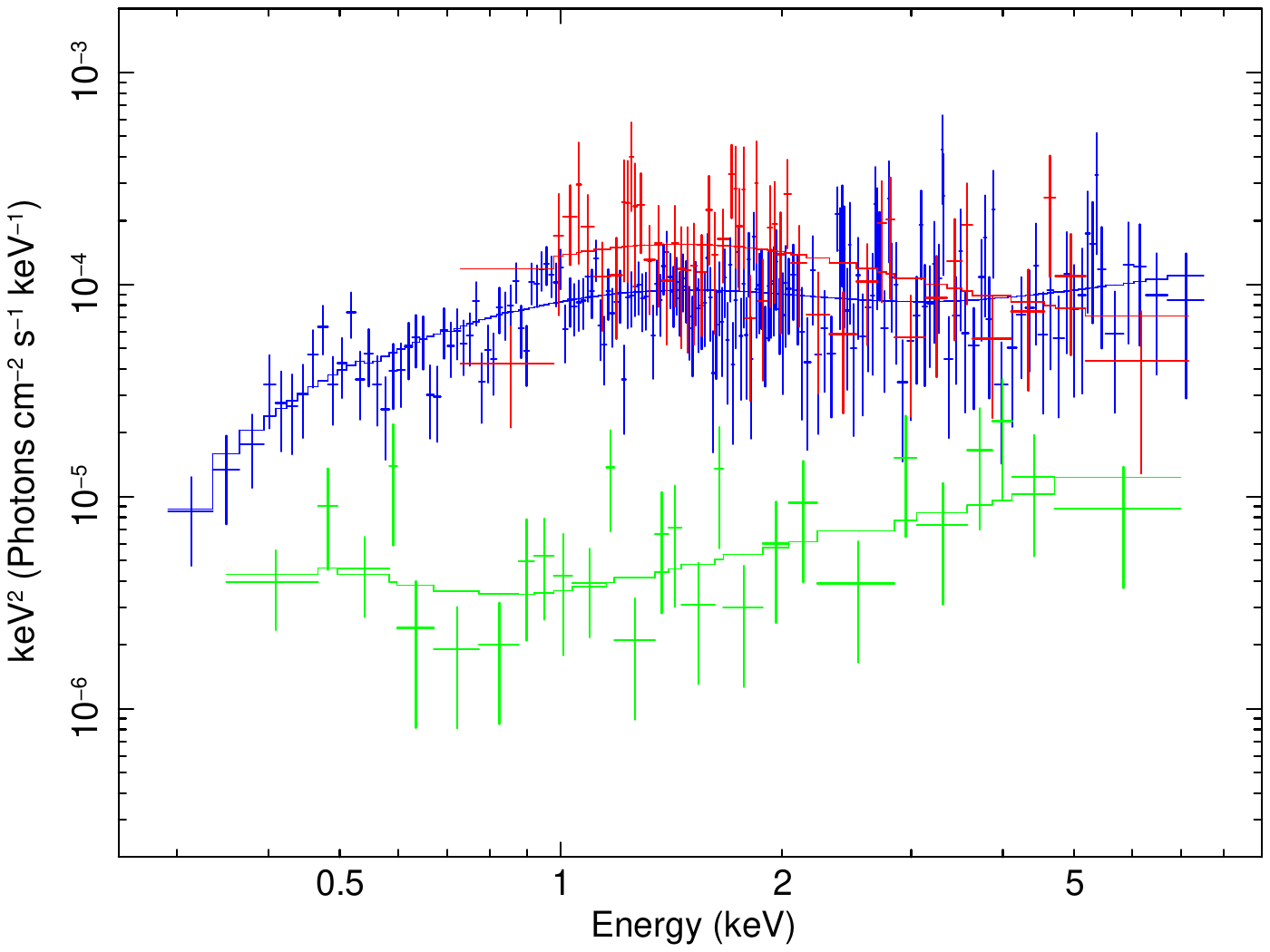}\\[-8pt]

\caption{Top panel: datapoints and data/model ratios for the {\it Chandra}/ACIS spectrum of X-2. Green is for ObsID 3495, blue for ObsID 3496 and red for ObsID 23572. The model is an absorbed Comptonization model ({\it tbabs} $\times$ {\it tbabs} $\times$ {\it simpl} $\times$ {\it diskbb}) (Table 2). The spectra were fitted with the Cash statistics. The datapoints were rebinned to a minimum signal to noise ratio of 2.7 for plotting purposes only.  Bottom panel: unfolded spectra corresponding to the same three observations.
}
  \label{x2_simpl_diskbb}
\end{figure}

\section{Main Results}

\begin{table}
\caption{Log of the {\it Chandra}, {\it HST}, VLT and ATCA observations used for this study.}
\vspace{-0.2cm}
\begin{center}
\begin{tabular}{lccc}  
\hline \hline\\[-5pt]    
\multicolumn{4}{c}{{\it Chandra}/ACIS}\\
\hline
ObsID & Obs.~Date &  Exp.~Time  & Aimpoint \\
 &   &  (ks) & \\
\hline  \\[-5pt]
3495 & 2002 May 21 & 23.4 & SN 2001ig \\
3496 & 2002 Jun 11 & 23.9 & SN 2001ig \\ 
23572 & 2020 Dec 02 & 5.0 &  SN 2001ig \\ 
\hline\\[-5pt] 
\multicolumn{4}{c}{{\it HST}}\\
\hline
Detector & Obs.~Date  &  Exp.~Time  &  Targets covered\\
   &    & (s)  &   \\
\hline  \\[-5pt]
WFPC2/F606W & 1994 Jul 16 & 160 & X-2\\
WFPC2/F814W & 2001 Jun 1 & 640  & X-2 \\ 
WFPC2/F450W & 2001 Jul 7 & 460 & R1\\
WFPC2/F814W & 2001 Jul 7 & 460  & R1 \\ 
WFC3/F275W & 2016 Apr 28 & 8694 & X-1, SN\,2001ig\\
WFC3/F336W & 2016 Apr 28 & 2920 & X-1, SN\,2001ig\\
\hline\\[-5pt]
\multicolumn{4}{c}{VLT/FORS2}\\
\hline
Grism & Obs.~Date  &  Obs.~Time &    Exp.~Time\\
   &    & (UT)  &   (s) \\
\hline  \\[-5pt]
 (Acq. image)  &  2011 Sep 01   & 02:05:26  &  30 \\
 300V &  2011 Sep 01   & 02:10:09  &  600 \\
 300V &  2011 Sep 01   & 02:20:52  &  600 \\
 (Acq. image) &  2011 Sep 01   & 02:35:20  &  30 \\
 1200R &  2011 Sep 01   & 02:40:48  &  900 \\
\hline\\[-5pt]

\multicolumn{4}{c}{ATCA}\\
\hline
Project ID & Obs.~Date  &  Exp.~Time  &  Bandwidth\\
   &    & (min)  & (GHz)  \\
\hline  \\[-5pt]
C3421 & 2021 Apr 10 & 370 & 2.0\\
C3421 & 2021 Apr 11 & 279  & 2.0 \\ 
C3421 & 2021 Apr 12 & 403 & 2.0 \\         
C3421 & 2021 Apr 13 & 195  & 2.0 \\ 

\hline\\[-5pt]
\end{tabular} 
\label{tab1}
\end{center}
\end{table}

\subsection{Astrometric alignment}
In addition to the use of new observations not available at the time, our multiband study improves on the results of \cite{soria06} because of a more accurate astrometric alignment between X-ray, optical and radio images. The two main reasons for this improvement are: i) an astrometric calibration based on the {\it Gaia} results \citep{gaia23}; ii) a new ATCA observation with higher sensitivity and more uniform coverage of the {\it uv} plane. 

In particular, the new 9-GHz ATCA images were not hampered by the side lobes that affected the 2001--2004 ATCA dataset, and  provided a more precise and accurate reference position of SN\,2001ig. Since this object is detected as a point-like source in {\it Chandra}, {\it HST}, Gemini and ATCA, it provides a useful anchor for relative astrometry between X-ray, optical and radio bands. We determined a radio position of SN\,2001ig of R.A.(J2000) $= 22^h\,57^m\,30^{s}.74 (\pm 0\farcs04)$, Dec.(J2000) $= -41^{\circ}\,02^{\prime}\,26\farcs35 (\pm 0\farcs05)$, which is $\approx$0\farcs7 away from the reference position currently listed in the SIMBAD database\footnote{\url{http://simbad.cds.unistra.fr/simbad/}} 
and in most SN catalogues in the literature. This revised position (unlike the old one) is perfectly consistent with the alignment of the {\it Chandra} and Gemini images onto the {\it Gaia} reference frame. This also enables a more precise localization of X-2 inside its associated star-forming complex and nebula (Figures 2--4), and a more fruitful search for the optical counterpart of X-1 (Section 3.8).

The first step was the alignment of the Gemini images. We selected eight {\it Gaia} sources with a bright, point-like, isolated, not saturated Gemini counterpart, as well as the radio position of SN\,2001ig. We calculated the average R.A. and Dec. offsets of those nine sources, and their scatter. We corrected the Gemini astrometry (simple translation) with standard {\sc iraf} {\it imcoords} tasks. After the re-alignment, we verified that the residual root-mean-square scatter of the Gemini positions with respect to the {\it Gaia} frame was $\approx$0\farcs1 in R.A. and $\approx$0\farcs1 in Dec. 

The second step was the alignment of the {\it HST}/WFPC2 and WFC3 images. For the WFPC2 fields that cover X-2 and the north-eastern sector of the galaxy, there are eight {\it Gaia} sources that appear point-like and unsaturated in the Wide Field chips of the WFPC2 F814W image, and six useful {\it Gaia}/WFPC2 associations in the F606W image. For the other pair of WFPC2 images, which cover R1 and the southern spiral arm, we found ten associations with {\it Gaia} sources.
For the WFC3/UVIS fields, we found seven useful associations in both the F275W and F336W images. For each {\it HST} image, we applied a simple coordinate translation to align to the {\it Gaia} frame. In addition, as a secondary calibrator, we checked the revised {\it HST} astrometry with the help of several other sources without a {\it Gaia} identification but with a bright point-like appearance in both Gemini and {\it HST}. We verified that after our astrometric improvement, there is no systemic offset between {\it HST} and Gemini positions (only random scatter within $\lesssim$0\farcs2).

\begin{table*}
\caption{Best-fitting parameters of the {\it Chandra}/ACIS-S spectra of X-2, fitted with the Cash statistics. Uncertainties for one interesting parameter are reported at the confidence interval of $\Delta C = \pm$2.70: this is asymptotically equivalent to the 90\% confidence interval in the $\chi^2$ statistics.
The Galactic absorption is fixed at $N_{\rm {H,Gal}} = 8.6 \times 10^{19}$ cm$^{-2}$.} 
\vspace{-0.3cm}
\begin{center}  
\begin{tabular}{lccc} 
 \hline 
\hline \\[-8pt]
  Model Parameters      &      \multicolumn{3}{c}{Values} \\
  & 2002 May 21       &       2002 June 11   &   2020 December 02  \\ %
\hline\\[-9pt]
   \hline\\[-9pt]
   \multicolumn{4}{c}{{\it tbabs} $\times$ {\it tbabs} $\times$ {\it simpl} $\times$ {\it diskbb}}\\
\hline\\[-6pt]
   $N_{\rm {H,int}}$   ($10^{22}$ cm$^{-2}$)   &  $ 0.03^{+0.04}_{-0.01}$ &     
   $ 0.03^{+0.04}_{-0.01}$  &  $ [0.03] $\\[4pt]
   $\Gamma$  &  $ 1.30^{+0.42}_{-0.16}$    &   $ 1.54^{+0.47}_{-0.45}$ & $ 2.38^{+1.34}_{-1.38} $  \\[4pt] 
   FracScatt   &   $ 0.61^{+0.23}_{-0.22}$      & $ 0.44^{+0.26}_{-0.31}$ & $ 0.40^{*}_{-0.26} $ \\[4pt]   
   $kT_{\rm {in}} $ (keV)
              &  $ 0.10^{+0.08}_{-0.05}$   &   $ 0.43^{+0.12}_{-0.13}$  & $ 0.46^{+0.27}_{-0.07} $\\[4pt]
   $N_{\rm {dbb}}$  (km$^2$)$^a$
              &  $ 25^{+110}_{-24}$   &   $ 0.46^{+1.21}_{-0.27}$  & $ 0.50^{4.9}_{-0.41} $\\[4pt]
    $R_{\rm {in}}\sqrt{\cos \theta} $  (km)$^b$ &  $ 6450^{+8520}_{-5360}$   &  $ 872^{+789}_{-311}$  & $ 909^{+696}_{-523}$\\[4pt]
   C-stat/dof     &      $71.0/67$ (1.06)   &     $239.7/263$ (0.91) &  $95.7/102$(0.91)\\[4pt]
   $f_{0.3-10}$ ($10 ^{-13}$ erg cm$^{-2}$ s$^{-1}$)$^c$ & $0.41^{+0.12}_{-0.11} $ & $ 4.37^{+1.10}_{-1.10}$ & $ 5.37^{+0.39}_{-0.90}$ \\[4pt]
   $L_{0.3-10}$ ($10 ^{39}$ erg s$^{-1}$)$^d$ & $ 0.62^{+0.18}_{-0.18}$ & $ 6.53^{+0.79}_{-0.71}$ & $ 8.03^{+2.08}_{-1.35}$  \\[4pt]

   \hline\\[-7pt]
\multicolumn{4}{c}{{\it tbabs} $\times$ {\it tbabs} $\times$ ({\it diskbb} + {\it powerlaw})}\\
\hline\\[-5pt]
   $N_{\rm {H,int}}$   ($10^{22}$ cm$^{-2}$)   &  $ 0.08^{+0.06}_{-0.05}$ &     $ 0.13^{+0.04}_{-0.03}$ & $ 0.67^{+0.45}_{-0.41}$\\[4pt]
   $\Gamma$   & $ 1.35^{+0.42}_{-0.48} $ & $ 2.09^{+0.13}_{-0.13} $ & $ 3.08^{+0.58}_{-0.53} $ \\[4pt]
   $N_{\rm {po}}^e$  & $ 3.90^{+1.45}_{-1.55} $ & 
   $ 10.29^{+1.125}_{-0.99} $ & $ 38.11^{+30.6}_{-16.38}$\\[4pt]
   $kT_{\rm in}$ (keV)     &  $ 0.09^{+0.10}_{-0.03}$       &   --  & --\\[4pt] 
   $N_{\rm {dbb}} $  (km$^2$)$^a$ &  $ 29^{+244}_{-28}$   &   -- & -- \\[4pt]
   $R_{\rm {in}}\sqrt{\cos \theta} $  (km)$^b$ &  $ 6930^{+14320}_{-6050}$   &  -- & -- \\[4pt]
   C-stat/dof     &      $70.3/67$ (1.05)     &    $242.2/265$ (0.91) & $93.6/103$ (0.91) \\[4pt]
   $f_{0.3-10}$ ($10^{-13}$ erg cm$^{-2}$ s$^{-1}$)$^c$ & $ 0.40^{+0.16}_{-0.11} $ & $ 4.17^{+0.3}_{-0.28}$ & $ 4.37^{+0.13}_{-0.74}$ \\[4pt]
   $L_{0.3-10}$ ($10^{39}$ erg s$^{-1}$)$^d$ &  $ 0.67^{+0.36}_{-0.17}$ & $ 6.68^{+1.54}_{-1.25}$ & $ 28.5^{+39.8}_{-13.9} $   \\[4pt]
      \hline\\[-7pt]
\multicolumn{4}{c}{{\it tbabs} $\times$ {\it tbabs} $\times$ ({\it apec} + {\it powerlaw})}\\
\hline\\[-5pt]
   $N_{\rm {H,int}}$   ($10^{22}$ cm$^{-2}$)  & $ 0.00^{+0.07}_{-0.00} $  & $ 0.08^{+0.03}_{-0.03} $ & $ 0.67^{+0.45}_{-0.41} $ \\[4pt]
   $kT_{\rm apec}$ (keV)    &   --  & $ 1.13^{+0.19}_{-0.15} $ & --   \\[4pt] 
   $N_{\rm {apec}}^f $  & -- & $ 2.31^{+1.48}_{-1.13} $ & -- \\[4pt] %
      $\Gamma$   & $1.52^{+0.35}_{-0.31}$& $ 1.88^{+0.16}_{-0.16} $ & $ 3.08^{+0.58}_{-0.54}$   \\[4pt]
   $N_{\rm {po}}^e$  & $ 0.43^{+0.13}_{-0.08}$ & $ 7.76^{+1.39}_{-1.22} $ & $ 38.11^{+30.6}_{-16.38}$ \\[4pt]
   C-stat/dof   & $73.8/69$ (1.07) & $229.0/263$ (0.87)  & $93.6/103$ (0.91)\\[4pt]
   $f_{0.3-10}$ ($10 ^{-13}$ erg cm$^{-2}$ s$^{-1}$)$^c$ & $ 0.31^{+0.10}_{-0.10}$ & $ 4.27^{+1.07}_{-1.07}$ & $ 4.37^{+0.13}_{-0.74}$\\[4pt]
   $L_{0.3-10}$ ($10 ^{39}$ erg s$^{-1}$)$^d$ & $ 0.50^{+0.14}_{-0.11}$ & $ 7.16^{+0.51}_{-0.48} $&
     $ 28.5^{+39.8}_{-13.9} $  \\[4pt]
\hline 
\vspace{-0.5cm}
\end{tabular}
\end{center}
\begin{flushleft} 
$^a$: $N_{\rm {dbb}} = (r_{\rm{in}}/d_{10})^2 \cos \theta$, where $r_{\rm {in}}$ is the apparent inner disk radius in km, $d_{10}$ the distance to the source in units of 10 kpc (here, $d_{10} = 1080$), and $\theta$ is our viewing angle ($\theta = 0$ is face-on). \\
$^b$: $R_{\rm {in}} \approx 1.19 r_{\rm in}$ for a standard disk \citep{kubota98}.\\ 
$^c$: observed fluxes in the 0.3--10 keV band\\
$^d$: isotropic unabsorbed luminosities in the 0.3--10 keV band, defined as $4\pi d^2$ times the de-absorbed fluxes.\\
$^e$: units of $10^{-5}$ photons keV$^{-1}$ cm$^{-2}$ s$^{-1}$ at 1 keV.\\
\end{flushleft}
\end{table*}

Third, we improved the {\it Chandra} astrometry, starting from the two longer observations. There are three associations of point-like X-ray sources with {\it Gaia} sources. A fourth reference point comes from the radio position of SN\,2001ig, which is also detected as an X-ray source in the 2002 {\it Chandra} observations. Finally, two {\it Chandra} sources have Gemini counterparts. We determined the observed centroids of the {\it Chandra} sources with the {\sc ciao} task {\it wavdetect}. Then, we corrected the {\it Chandra} coordinates with the tasks {\it wcs\_match} and {\it wcs\_update}. We did so first using only the three {\it Gaia} and one ATCA associations; then, including also the two Gemini associations. We obtained the same result. We also verified that there is no need to include rotation and scaling corrections, as they do not improve the alignment compared with a simple translation. The third, shorter {\it Chandra} observation was aligned to the coordinates of the first two observations, based on the brightest X-ray sources visible in all three looks. For the main target of our study, the X-2 ULX, we obtain a refined position of R.A.(J2000) $= 22^h\,57^m\,24^{s}.71 (\pm 0\farcs2)$, Dec.(J2000) $= -41^{\circ}\,03^{\prime}\,44\farcs1 (\pm 0\farcs2)$. 

Taking into account the residual random scatter in the X-ray and optical positions (combined in quadrature), we estimate that we can locate the position of the brightest X-ray sources onto the {\it HST}/WFPC2 and Gemini images within a 90\% confidence radius of $\approx$0\farcs4, and onto the {\it HST}/WFC3 images within a 90\% confidence radius of $\approx$0\farcs3. The greater precision of the UVIS alignment is partly due to their smaller pixel size (0\farcs04, compared with the resampled pixel size of 0\farcs1 for the WFPC2 images), and partly to the fact that the UVIS field include the optical counterpart of SN 2001ig, whose radio coordinates are now well-determined from the ATCA maps.

\subsection{X-ray properties of X-2}
In 2002, X-2 rose from $L_{\rm X} \approx 6 \times 10^{38}$ erg s$^{-1}$ to $L_{\rm X} \approx 6 \times 10^{39}$ erg s$^{-1}$ in the space of 20 days (Figure 5, Table 2 and \citealt{soria06}). In the 2020 {\it Chandra} data, it is seen again in an ultraluminous state. We fitted the new spectrum with the C statistics, because of the limited number of counts (caused by a short exposure time and a sharp decline in ACIS sensitivity since 2002). For consistency, we also refitted the two spectra from 2002 with the C statistics, which for a high number of counts give identical results to the $\chi^2$ fitting used in \cite{soria06}.
 
The 2020 spectrum is mildly curved (Figure 5), well fitted (C-stat $= 95.7/102$ degrees of freedom) by standard Comptonization models; {\it e.g.}, {\it simpl} $\times$ {\it diskbb} (Table 2), based on the Comptonization model of \cite{steiner09} applied to a seed disk-blackbody spectrum. For this fit, we froze the value of the intrinsic column density in the 2020 dataset to the best-fitting value obtained in the 2002 spectra ($N_{\rm H} \approx 3 \times 10^{20}$ cm$^{-2}$), because such low values of $N_{\rm H}$ are essentially unconstrained with the current low sensitivity of ACIS below 0.8 keV. The unabsorbed isotropic 0.3--10 keV luminosity for the Comptonization model is $L_{\rm X} \approx 8^{+2}_{-1} \times 10^{39}$ erg s$^{-1}$. A simple power-law model (Table 2) can fit the 2020 data equally well (C-stat $= 93.6/103$) but only with the addition of a suspiciously high level of intrinsic absorption ($N_{\rm H} \approx 7 \times 10^{21}$ cm$^{-2}$), inconsistent with the low values of $N_{\rm H}$ found in the much deeper 2002 observations. An absorbed {\it diskbb} model provides a worse fit (C-stat $= 101.3/103$) than a power law or a Comptonization model, because it has too much spectral curvature. A $p$-free disk gives a marginal improvement, (C-stat $= 98.6/102$) for a characteristic temperature $kT_{\rm in} \approx 1.2$ keV and $R_{\rm {in}}\sqrt{\cos \theta} \approx 80$ km. The unabsorbed 0.3--10 keV luminosity of the {\it diskbb} and {\it diskpbb} model fits are $L_{\rm X} \approx (9\pm2) \times 10^{39}$ erg s$^{-1}$. 

For the 2002 June 11 dataset, we confirm (in agreement with what was reported by \citealt{soria06}) that the fit is significantly improved with the addition of a thermal plasma component (modelled in {\sc xspec} with {\it apec}\footnote{\url{https://heasarc.gsfc.nasa.gov/xanadu/xspec/manual/node134.html}}) to any smooth continuum model. For example (Table 2), adding the optically thin thermal plasma emission ($kT_{\rm apec} \approx (1.1 \pm 0.2)$ keV) to an absorbed power-law model brings the C statistics from 242.5 over 265 degrees of freedom, down to 229.0 over 263 degrees of freedom: an improvement significant to the 99\% confidence limit ({\it simftest} in {\sc xspec}). Similar improvements of $|\Delta C| \approx 13$ are also obtained when an {\it apec} component is added to a Comptonization model. The presence of emission line residuals in the 2002 June 11 spectrum, at high $L_{\rm X}$, can be interpreted as typical evidence of ULX outflows \citep{Middleton2015,Pinto2016,kosec18}. In 2020, the X-ray luminosity was even higher, but unfortunately the exposure time was shorter and ACIS has lost too much sensitivity below 1 keV (Figure 5), so that we cannot test the significance of an additional {\it apec} component in that epoch.

In summary, the 2020 {\it Chandra} observations show that the X-2 ULX remains very bright--in fact, more luminous than in 2002. This supports the hypothesis that X-2 provides an abundant supply for ionizing photons for the surrounding nebula. Extrapolating the best-fitting Comptonization model {\it simpl} $\times$ {\it diskbb} for the 2020 dataset, we estimate an intrinsic flux of $\approx$3.5 $\times 10^{48}$ photons s$^{-1}$ emitted in the 54--300 eV range. For the best-fitting {\it diskpbb} model, the isotropic flux is $\approx$2 $\times 10^{49}$ photons s$^{-1}$ in the same energy range. The average luminosity and ioinizing flux over all three {\it Chandra} observations is of course lower, because of the lower state seen in 2002 May 21 (Table 2).

\begin{figure}
\centering
\includegraphics[width=\columnwidth]{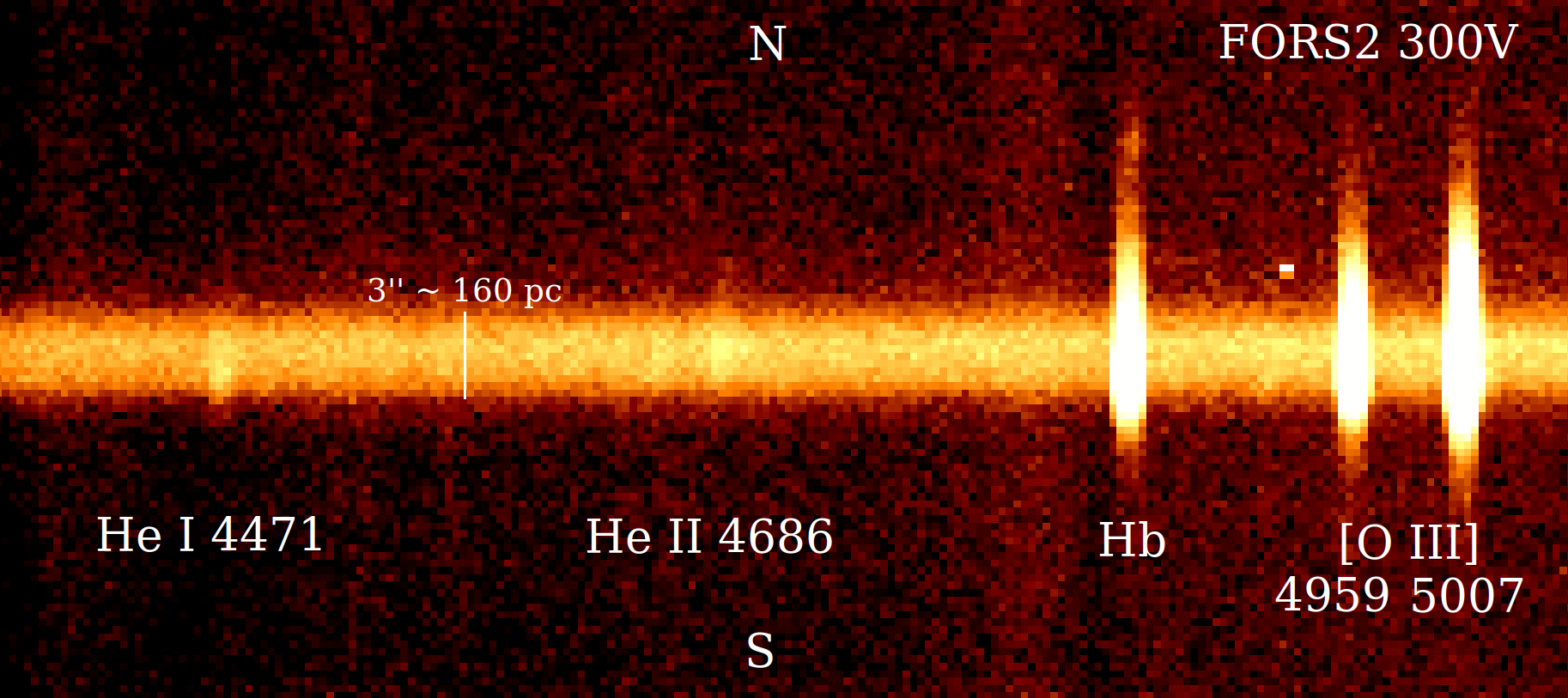}\\
\includegraphics[width=\columnwidth]{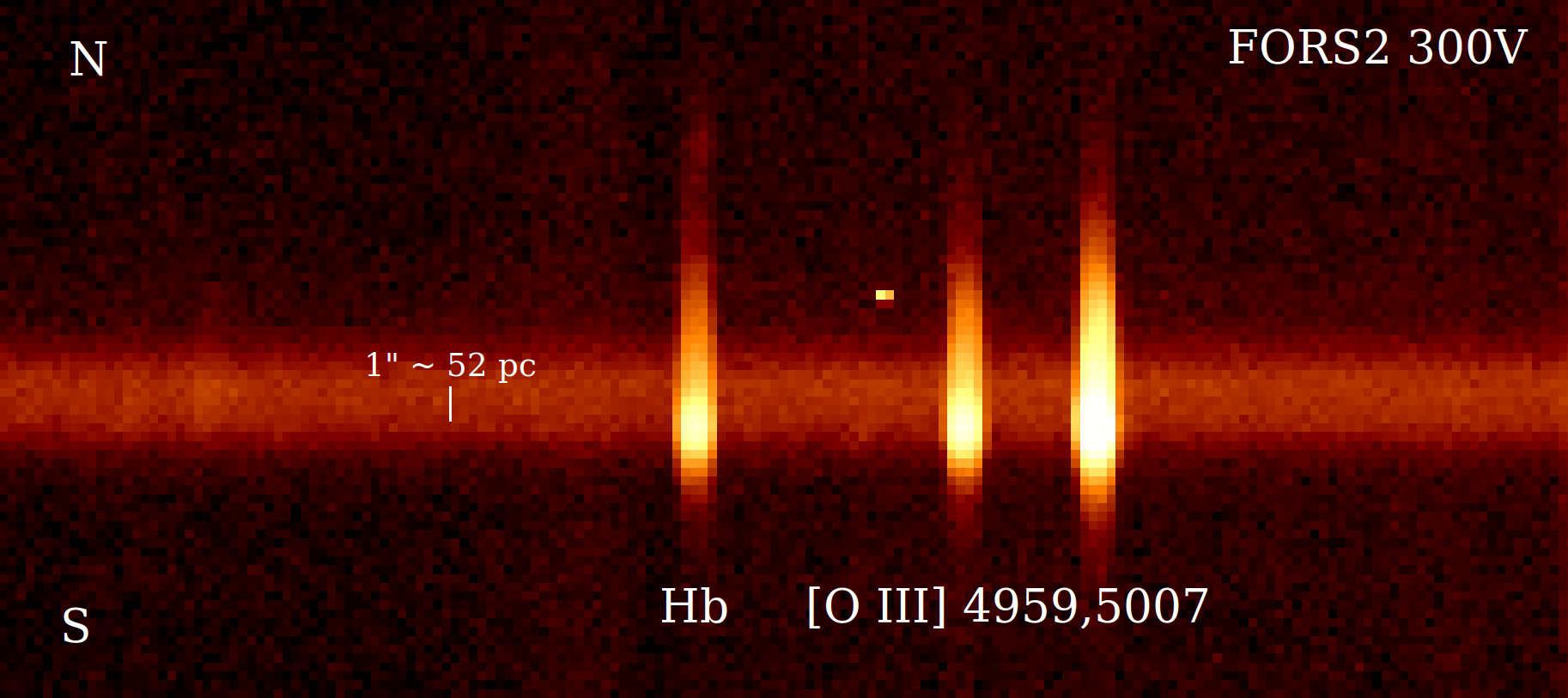}
 \caption{Top panel: portion of the VLT/FORS2 2D spectrum of the nebula and star-forming region around X-2, taken with VLT/FORS2, 300V grism. The Y axis represents the spatial direction (slit running from north to south), with a dispersion of \protect{0\farcs25} per pixel.  The X axis is the wavelength direction, with a spectral dispersion of 3.3 \AA\ per pixel and instrumental resolution of 9.4 \AA\ FWHM. The image highlights the different location of the lower-ionization lines emission (closer to the southern part of the star-forming region) compared with the \ion{He}{II} $\lambda$ 4686 emission (closer to the northern part). 
 Bottom panel: zoomed-in view of the same spectrum, around the H$\beta$-[\ion{O}{III}] complex, displayed with the different color scale, to show that all three lines peak on the southern end of the star-forming region. The vertical marker (4 pixels $\approx$ 1\farcs0) represents the spacial displacement between the peak of the \ion{He}{II} $\lambda$ 4686 line and the peak of the other lines. 
 }
  \label{fors2_blue}
\end{figure}

\begin{figure}
\centering
\includegraphics[width=\columnwidth]{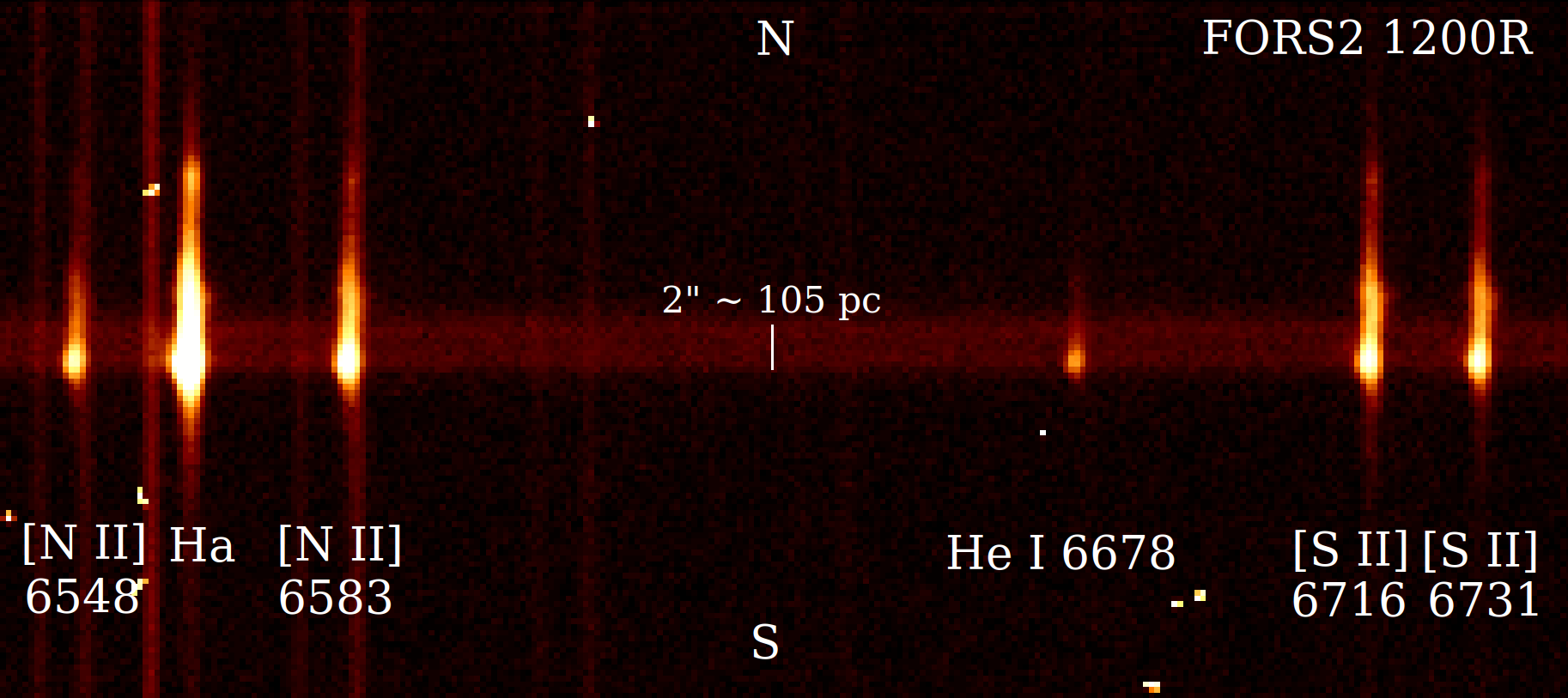}\\
\includegraphics[width=\columnwidth]{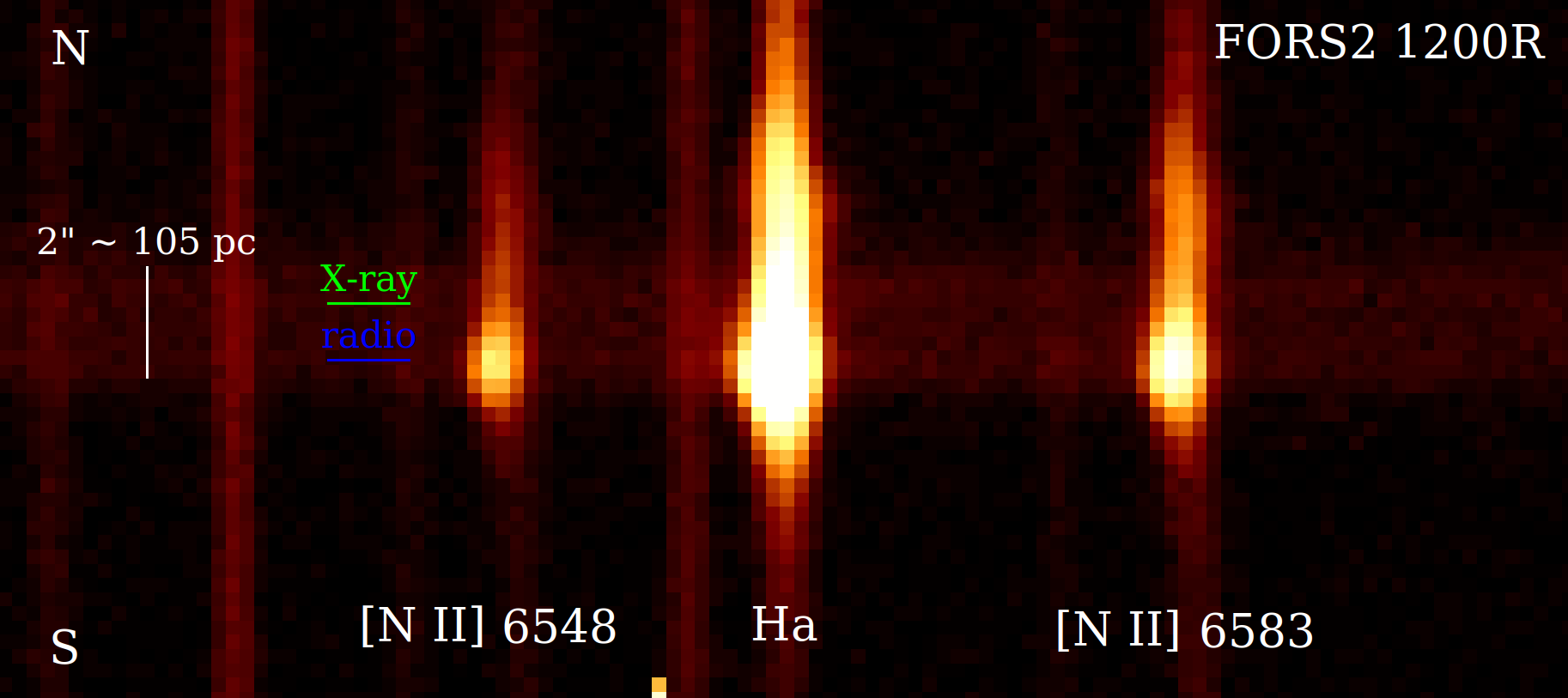}
 \caption{Top panel: portion of the VLT/FORS2 2D spectrum of the nebula and star-forming region around X-2, taken with VLT/FORS2, 1200R grating. The Y axis represents the spatial direction from north to south, with a dispersion of 0\farcs25 per pixel.  The X axis is the wavelength direction, with a spectral dispersion of 0.75 \AA\ per pixel, and a resolution of 1.8 \AA\ FWHM. Bottom panel: zoomed-in view around the H$\alpha$-[\ion{N}{II}] complex. The green marker indicates the spatial location of the X-ray source (corresponding to the peak \ion{He}{II} $\lambda$ 4686 emission); the blue marker corresponds to the location of the unresolved radio source.}
  \label{fors2_red}
\end{figure}

\subsection{Morphology of the star-forming complex around X-2}
The star-forming region around X-2 has a central region (diameters of $\approx$100 $\times$ 150 pc) with bright continuum emission from young stars and star clusters (Gemini and {\it HST} images, Figures 2,3) together with strong line emission. In addition, line-emitting gas without significant broadband emission extends at least another 200 pc to the north and 100 pc to the south of the starlight-dominated region (compare the north-south extent of continuum and line emission in Figures 6,7).

In the continuum-dominated regions, two specific location stand out for their multiband properties. Near the southern part of the complex, we find the brightest star cluster, coincident (within a 90\% confidence limit of 0\farcs2) with the unresolved radio source (Figures 3,4). About 1\farcs1 north of the radio source, in the central/upper part of the star-forming complex, it is the location of the X-2 ULX. There is at least one point-like optical source (a candidate star cluster) within the {\it Chandra} error circle, which must be considered the best candidate optical counterpart of the ULX, but nothing outstanding compared with other young star clusters in the surroundings. It was already noted by \cite{soria06} that X-ray and radio positions appeared to be different; now we can confirm that result, with the help of the 2021 ATCA observations and a more accurate multi-instrument astrometric alignment. Moreover, we have now ascertained that the radio source coincides with the optically brightest star cluster.

For the radio source, the 2021 ATCA data at 9 GHz give a centroid at R.A.(J2000) $= 22^h\,57^m\,24^{s}.697 (\pm 0\farcs07)$, Dec.(J2000) $= -41^{\circ}\,03^{\prime}\,45\farcs22 (\pm 0\farcs13)$. The flux density is $F_{\rm 5.5GHz} = (135 \pm 12) \mu$Jy at 5.5 GHz (corresponding to a luminosity density $L_{\rm 5.5GHz} = (1.9 \pm 0.2) \times 10^{25}$ erg s$^{-1}$ Hz$^{-1}$) and 
$F_{\rm 9GHz} = (70 \pm 11) \mu$Jy at 9.0 GHz ($L_{\rm 9.0GHz} = (9.0 \pm 2.0) \times 10^{24}$ erg s$^{-1}$ Hz$^{-1}$). These values are consistent with the average flux densities determined from the 2001--2004 dataset, namely  $F_{\rm 4.8GHz} = (138 \pm 34) \mu$Jy at 4.8 GHz and 
$F_{\rm 8.6GHz} = (100 \pm 45) \mu$Jy at 8.6 GHz  \citep{soria06}. The spectral index $\alpha = -(1.4 \pm 0.4)$ suggests optically thin synchrotron emission, typical of SN remnants and/or ULX bubbles.

The radio luminosity $\nu L_{\nu} \approx 1.0 \times 10^{35}$ erg s$^{-1}$ ({\it i.e.}, $\approx$2.5 times more luminous than Cas A) is near the upper end of the observed radio luminosity function for SNRs in nearby galaxies \citep{chomiuk09,thompson09}. Models of SNR radio luminosity evolution suggest that such high values can be reached for a normal SN (explosion energy $\approx$10$^{51}$ erg) exploding in an relatively dense ambient medium with $n_e \gtrsim 10$ cm$^{-3}$ \citep{berezhko04,sarbadhicary17,sarbadhicary19,leahy22}. We will see from our diagnostic line ratio analysis (Section 3.6) that such higher-than-usual interstellar medium (ISM) density is indeed plausible.

The bright star cluster coincident with the radio source is the peak both of the continuum and of the Balmer line emission. From the {\it HST}/WFPC2 images, we obtain an apparent brightness of $m_{606W} = (19.9\pm 0.1)$ mag (Vega system) and $m_{814W} = (20.2\pm 0.1)$ mag. If we only consider line-of-sight Galactic extinction, this corresponds to dereddened absolute magnitudes $M_{606W} = -(10.3\pm 0.1)$ mag and $M_{814W} = -(10.0\pm 0.1) $ mag. However, our analysis of the Balmer decrement in the VLT spectra suggest a higher intrinsic reddening, as we shall discuss later (Section 3.4).
Moving now our attention to the upper part of the star-forming complex, near the location of the X-ray source, we see in the {\it HST}/WFPC2 images several optical peaks, candidate (small) star clusters, standing out from an unresolved, bright background (Figures 3,4). Only one of those sources is fully inside the {\it Chandra} error circle, and is consistent with the peak of the \ion{He}{II} $\lambda$4686 emission (Section 3.6).  
The apparent brightness of that source is $m_{606W} = (22.3 \pm 0.1)$ mag, $m_{814W} = (21.9 \pm 0.1)$ mag. If corrected only for line-of-sight Galactic extinction, this corresponds to de-reddened absolute brightness $M_{606W} = -(7.9\pm 0.1)$ mag, $M_{814W} = -(8.3\pm 0.1)$ mag. However, we will discuss how also in this case, the Balmer decrement measured in the VLT spectra suggests additional intrinsic reddening. 


None of the individual optical sources in the northern part of the bright complex (inside or at the edge of the {\it Chandra} error circle for the ULX) is spatially resolved. From the broadband continuum brightness and colours alone, we cannot completely rule out that they are individual supergiant stars rather than small star clusters. For example, using the Padova isochrones computed with the {\sc parsec} code\footnote{\url{http://stev.oapd.inaf.it/cgi-bin/cmd_3.7}} \citep{bressan12}, we find that the optical source closest to the {\it Chandra} position is also consistent with a yellow supergiant with initial mass in the range of $\approx$17--20 $M_{\odot}$ at an age of $\approx$9.0--11.5 Myr, and with a radius of $\approx$240--330 $R_{\odot}$. However, we consider this scenario very unlikely, because the strong Balmer line emission (Section 3.4) indicates a much younger age for the whole star-forming clump, an age at which stars in this mass range have not left the main sequence yet. 





\begin{table}
\caption{Observed fluxes (relative to H$\beta \equiv 1.00$) of the main lines detected in our VLT/FORS2 spectra around the position of the X-2 ULX and around the position of the brightest star cluster (SC) and radio source ($\approx$1$^{\prime\prime}$ south of the ULX). Spatial extent of the extraction region: $\pm$0\farcs5 along the slit in the north-south direction, around each of the two positions.} 
\begin{center}  
\begin{tabular}{lccc} 
 \hline 
\hline \\[-8pt]
  Line ID      &   $\lambda_{\rm obs}$  & Flux near X-2 & Flux near SC\\
  & (\AA)       &     &    \\ %
    \hline\\[-7pt]
[O {\footnotesize {II}}]$\lambda\lambda$3726,3729 & 3739.3 &  2.76  & 2.20 \\
H10 $\lambda 3798$ & 3810.0  & $<$0.02 & 0.039  \\
H9 $\lambda 3835$ & 3846.5  & 0.025 & 0.053 \\

[Ne {\footnotesize {III}}]$\lambda$3869  & 3880.7   &   0.29  & 0.22 \\   
H8 $\lambda 3889$ $+$ He {\footnotesize {I}} $\lambda 3889$ &   3900.8 & 0.16  &  0.15 \\
H$\epsilon$ $\lambda 3970$ $+$ [Ne {\footnotesize {III}}]$\lambda 3967$  & 3981.2  &  0.18  & 0.17 \\   
H$\delta$ $\lambda 4102$  & 4114.2  &   0.19 & 0.18 \\     
H$\gamma$ $\lambda 4340$  & 4353.2  &   0.49  & 0.46\\  

[O {\footnotesize {III}}] $\lambda 4363$  & 4376.5  &  0.047 & 0.031 \\    
He {\footnotesize {I}} $\lambda 4471$   & 4484.8    &  0.047  & 0.047 \\   
$^{a,b}$He {\footnotesize {II}} $\lambda 4686$   & 4698.9   &  0.073  & 0.013 \\
$^{c,d}$H$\beta$ $\lambda 4861$  &   4875.9   &  1.00 & 1.00 \\

[O {\footnotesize {III}}] $\lambda 4959$ & 4973.4   &   1.22   & 1.21  \\

[O {\footnotesize {III}}] $\lambda 5007$ & 5021.4   &  3.67   &  3.65 \\

[N {\footnotesize {I}}] $\lambda 5198$  & 5213.5  & 0.010  & 0.015    \\
He {\footnotesize {I}} $\lambda 5876$  &  5893.2  & 0.15 & 0.15 \\

[O {\footnotesize {I}}] $\lambda 6300$   & 6319.2 & 0.068 & 0.045 \\

[S {\footnotesize {II}}] $\lambda 6313$  & 6331.0  & 0.026 &  0.021\\ 

[O {\footnotesize {I}}] $\lambda 6364$ & 6382.8  & 0.027 & 0.015 \\

[N {\footnotesize {II}}] $\lambda 6548$  & 6567.8  &  0.18  &  0.18  \\
$^{e,f}$H$\alpha$ $\lambda 6563$ & 6582.5  & 3.54 & 4.08 \\

[N {\footnotesize {II}}] $\lambda 6583$   & 6603.2  & 0.55  & 0.56 \\

[N {\footnotesize {II}}] $\lambda 6596$  & 6615.6  & $<$0.005  & 0.005    \\
He {\footnotesize {I}} $\lambda 6678$  & 6698.1  & 0.049 & 0.054\\

[S {\footnotesize {II}}] $\lambda 6716$  & 6736.5   &  0.41  & 0.32 \\   

[S {\footnotesize {II}}] $\lambda 6731$ & 6751.0    &  0.30  & 0.26\\    
He {\footnotesize {I}} $\lambda 7065$ & 7086.2  & 0.037  & 0.052\\ 

[Ar {\footnotesize {III}}] $\lambda 7136$ & 7157.1   &  0.14 & 0.19 \\  

[N {\footnotesize {II}}] $\lambda 7215$  & 7236.5  &  0.030 & 0.030  \\

[O {\footnotesize {II}}]$\lambda\lambda$7320,7330 & 7345.0 &  0.076  & 0.092\\

[Ar {\footnotesize {III}}] $\lambda 7751$ & 7775.4   & 0.042  & 0.045 \\  
\hline 
\vspace{-0.5cm}
\end{tabular}
\end{center}
\begin{flushleft} 
$^a$: Observed fluxes of He {\footnotesize {II}} $\lambda 4686$: $F(4686) = (0.10\pm0.01) \times 10^{-15}$ erg cm$^{-2}$ s$^{-1}$ on the 1$^{\prime\prime}$ slit around X-2; $F(4686) = (0.04\pm0.01) \times 10^{-15}$ erg cm$^{-2}$ s$^{-1}$ on the 1$^{\prime\prime}$ slit around the brightest SC; $F(4686)  \approx 0.3 \times 10^{-15}$ erg cm$^{-2}$ s$^{-1}$ extrapolated from the whole continuum-emitting star-forming clumps;\\
$^b$: EW$(4686) = (5.3\pm0.5)$ \AA\ around X-2; EW$(4686) = (2.8\pm0.5)$ \AA\ around the brightest SC;\\
$^c$: Observed fluxes of H$\beta$: $F({\mathrm{H}}\beta) = (1.4\pm0.1) \times 10^{-15}$ erg cm$^{-2}$ s$^{-1}$ on the 1$^{\prime\prime}$ slit around X-2; $F({\mathrm{H}}\beta) = (3.0\pm0.1) \times 10^{-15}$ erg cm$^{-2}$ s$^{-1}$ on the 1$^{\prime\prime}$ slit around the brightest SC; $F({\mathrm{H}}\beta)  \approx (9\pm1) \times 10^{-15}$ erg cm$^{-2}$ s$^{-1}$ extrapolated from the whole continuum-emitting star-forming clumps;\\
$^d$: EW$({\mathrm{H}}\beta) = (80\pm5)$ \AA\ around X-2; EW$({\mathrm{H}}\beta) = (230\pm10)$ \AA\ around the brightest SC;\\
$^e$: Observed fluxes of H$\alpha$: $F({\mathrm{H}}\alpha) = (5.0\pm0.1) \times 10^{-15}$ erg cm$^{-2}$ s$^{-1}$ on the 1$^{\prime\prime}$ slit around X-2; $F({\mathrm{H}}\alpha)  = (12.3\pm0.1) \times 10^{-15}$ erg cm$^{-2}$ s$^{-1}$ on the 1$^{\prime\prime}$ slit around the brightest SC; $F({\mathrm{H}}\alpha)  \approx (37\pm5) \times 10^{-15}$ erg cm$^{-2}$ s$^{-1}$ extrapolated from the whole continuum-emitting star-forming clumps;\\
$^f$: EW$({\mathrm{H}}\alpha) = (400\pm20)$ \AA\ around X-2; EW$({\mathrm{H}}\alpha) = (1350\pm50)$ \AA\ around the brightest SC.\\

\end{flushleft}
\end{table}


\subsection{Star formation properties near X-2 from the VLT spectra}

For a better understanding of the star formation properties of this clump, we turn to the VLT/FORS2 spectra. Following up on the arguments discussed in Section 3.3, we identified two characteristic regions: one around the southern star cluster and radio source, and the other around the {\it Chandra} source, $\approx$1$^{\prime\prime}$ to the north. Thus, we defined two extraction regions along the slit (both for the 300V and 1200R grisms), centred on the two positions, and with a spatial extent of 4 pixels $\approx$1$^{\prime\prime}$. In Table 3, we summarize the main properties of the lines detected in the two regions. This is more informative than the average properties over the whole complex, as we are looking for differences between the two positions.

We do not find significant velocity differences between northern and southern section in any of the lines. Thus, in Table 3 we report the average central wavelength of each line measured from the spectrum of the whole complex. The average recession speed is of $(890 \pm 10)$ km s$^{-1}$. This redshift is in excellent agreement with the expected value at the location of X-2, derived from the neutral hydrogen maps of \cite{sorgho19} and \cite{sardone21}. 

We selected the five strongest lines in the 1200R spectra to constrain line broadening, comparing their observed FWHMs from the X-2 region with those of sky lines. We obtain a marginally significant result of a mean intrinsic FWHM $\approx 0.9$ \AA\ around the X-ray source position and intrinsic FWHM $\approx 0.8$ \AA\ around the radio source position. Thus, the intrinsic FWHM is $\lesssim$ 40 km s$^{-1}$, a plausible value considering thermal broadening and turbulent motion of the gas in an \ion{H}{II} region. Moreover, we did double Gaussian fits of the strongest lines to test for the presence of a secondary, broader component in addition to the main narrow component. We found that the line profile in the radio source region does hint at the presence of broader wings extending up to $\pm$120 km s$^{-1}$ from the central position. This is consistent with a shock ionization component associated with that non-thermal radio emitting region. The flux in the broader component is only a few percent of the total line flux. We would need higher dispersion spectra to investigate that further. No such broad component was found for the region around the x-ray source. We also did not find any evidence of P-Cygni profiles (signature of strong outflows) in the Balmer lines.


The flux of each line was computed from the average of the two 300V spectra, whenever possible. This was done for two reasons. First, because the 300V grisms covers both the red and the blue part of the optical spectrum, which reduces potential systematic errors for a flux comparison, for example between H$\alpha$ and H$\beta$. Second, because the 1200R spectrum was taken in not perfectly photometric conditions (thin cirrus clouds). However, some of the lines can only be resolved in the 1200R spectrum: for example, [O {\footnotesize{I}}] $\lambda$6300 and [S {\footnotesize{II}}] $\lambda$6313 are partly blended in the 300V spectra, and so are [N {\footnotesize {II}}] $\lambda 6548$ and H$\alpha$. In that case, we measured the relative flux of those lines to H$\alpha$ in the 1200R spectrum, and then converted those values to absolute fluxes based on the 300V measurements.

In this Section, and in Section 3.5, we focus on the properties of the Balmer lines. Properties of other lines will be discussed in Sections 3.6 and 3.7.
For very young star clusters, the observed H$\alpha$ and H$\beta$ EWs and flux ratio (Balmer decrement) are a much better proxy for age and intrinsic reddening than the broadband continuum. From the VLT spectra, we determine EW(H$\alpha$) $\approx (1350\pm50)$\AA, and EW(H$\beta$) $\approx (230\pm10)$ \AA\ at the location of the southern optical/radio peak (Table 4). Star cluster simulations with {\sc starburst99}\footnote{\url{https://www.stsci.edu/science/starburst99/docs/default.htm}} \citep{leitherer99,leitherer14}, assuming instantaneous star formation and solar metallicity, indicate a well-constrained age of $\approx (2.5 \pm 0.2)$ Myr for the observed EWs. The Balmer ratio $F({\mathrm H}\alpha)/F({\mathrm H}\beta) \approx 4.08$ is significantly higher than the canonical value of 2.86 for photoionized gas. Using a Milky Way extinction curve with $A_V = 3.1 E(B-V)$, the total (intrinsic plus line-of-sight Galactic) reddening required to explain the high Balmer ratio is $E(B-V) \approx 0.36$ mag, corresponding to $A_V \approx 1.11$ mag, $A_{606W} \approx 1.00$ mag, $A_{814W} \approx 0.66$ mag. Subtracting this extinction from the observed values of optical brightness (Section 3.3), we obtain our best estimate for the de-reddened absolute brightness $M_{606W} = -(11.3\pm 0.1)$ mag, $M_{814W} = -(10.6\pm 0.1)$ mag. This agrees well with the expected broadband colour between the two bands for a star cluster at the age of $\approx$2.5 Myr\footnote{{\sc starburst99} predicts $-0.04 \lesssim V-I ({\rm mag}) \lesssim 0.1$ for a cluster age of $\approx (2.5 \pm 0.2)$ Myr (instantaneous star formation, solar metallicity, Kroupa IMF). The F814W filter is essentially identical to the standard $I$ band. Instead, the F606W filter is broader than the standard $V$ towards longer wavelengths, and includes H$\alpha$ \citep{holtzman95}. For a young star cluster with EW(H$\alpha$) $\approx 1350$ \AA, we estimate that $m_{606W}$ is $\approx$0.6 mag brighter than standard $V$. Thus, we estimate that $M_{606W} \approx -11.3$ mag corresponds to $M_{V} \approx -10.7$ mag, $V-I \approx (-0.1\pm0.1)$ mag.}.
Combining the cluster age and the extinction-corrected $I$-band absolute brightness of $M_{814W} \approx -10.6$ mag, from {\sc starburst99} we obtain an estimate of $(20,000\pm2000) M_{\odot}$ for the total stellar mass in the southern cluster.

We can apply similar arguments for the smaller star clusters near the {\it Chandra} position. The VLT spectra extracted from $\pm$0\farcs5 around the ULX position give EW(H$\alpha$) $\approx (400\pm20)$\AA, and EW(H$\beta$) $\approx (80\pm5)$ \AA\ (Table 4), and a Balmer ratio $F({\mathrm H}\alpha)/F({\mathrm H}\beta) \approx 3.54$.
The EWs suggest an age of $\approx (4.4 \pm 0.3)$ Myr for the central/northern part of the clump, consistent with the scenario that star formation is propagating southwards in time.  The Balmer decrement implies a total reddening $E(B-V) \approx 0.22$ mag, which corresponds to $A_V \approx 0.67$ mag, $A_{606W} \approx 0.60$ mag, $A_{814W} \approx 0.40$ mag. If we apply these values of age and extinction to the point-like optical source closest to the X-ray position, we find absolute brightnesses $M_{606W} = -(8.5\pm 0.1)$ mag, $M_{814W} = -(8.7\pm 0.1)$ mag, which correspond to a stellar mass of $(2400\pm300) M_{\odot}$, and similar masses $\sim$2,000--3,000 $M_{\odot}$ for each of the other three small clusters within $\approx$0\farcs5 of the X-ray position (Figure 3).

An average extinction $A_V \approx 0.67$ mag corresponds to an equivalent neutral hydrogen column density $N_{\rm H} \sim 1$--2 $\times 10^{21}$ cm$^{-2}$, depending on the choice of conversion factor in the literature \citep{foight16,willingale13,watson11,guver09}. This is marginally higher than the total values of $N_{\rm H} \lesssim 10^{21}$ cm$^{-2}$ fitted to the {\it Chandra} spectra in the most reliable Comptonization or disk-blackbody models (Table 2). However, the optical extinction is an average value over a projected $\approx$50 pc $\times$ 50 pc region around the X-ray source. The latter may be in a lower-density cavity (perhaps because the gas is ionized by the ULX itself), or may be located on the near side of the ionized nebula rather than inside one of the small star clusters.


\begin{figure}
\vspace{-0.7cm}
\hspace{-0.4cm}
\includegraphics[height=1.25\columnwidth, angle=270]{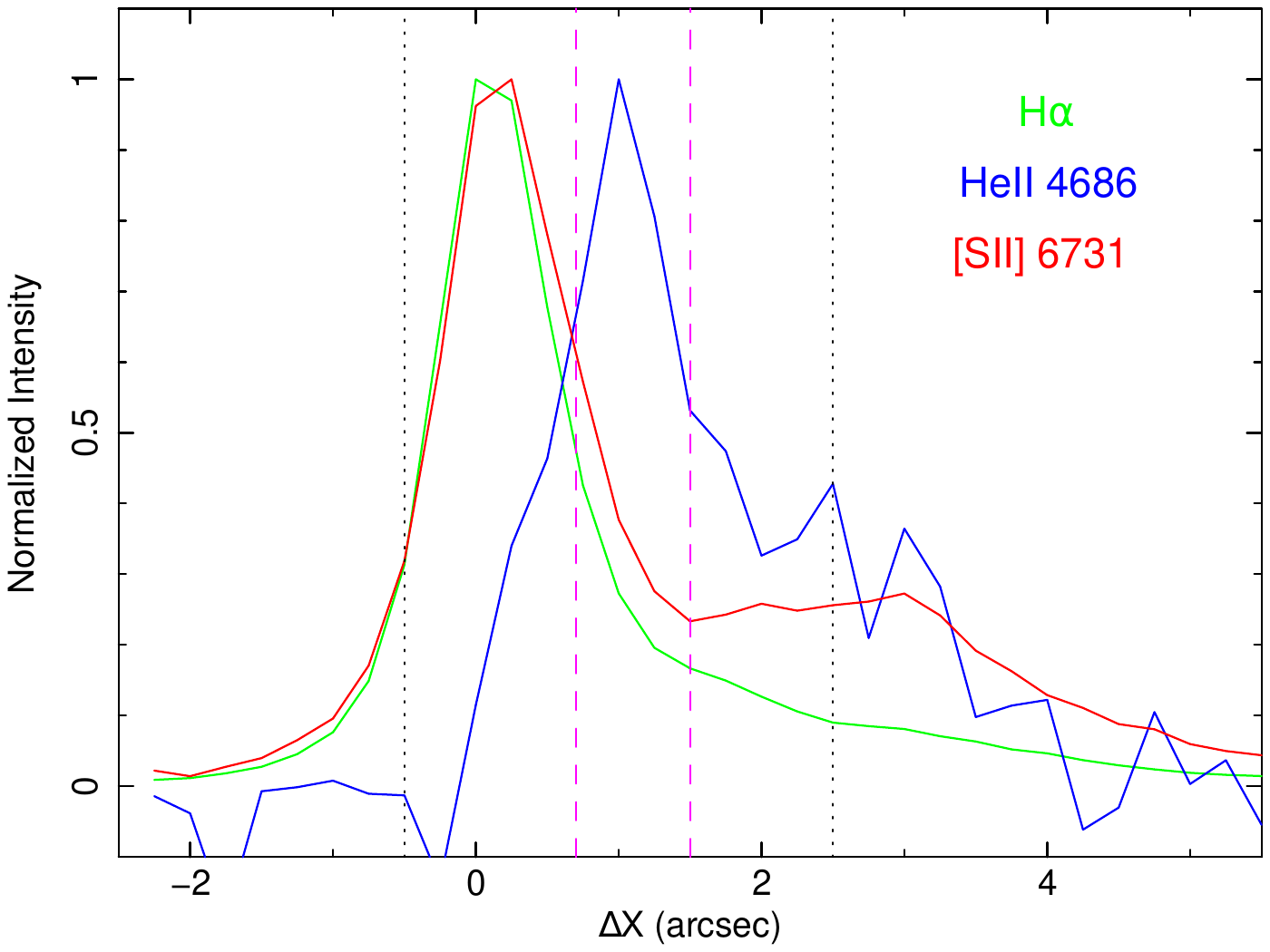}\\[-8pt]
 \caption{Spatial profile of a few diagnostic lines, along the slit in the north-south direction (south is left and north is right, along the X axis). The reference position ($\Delta X= 0$) on the X axis corresponds to the peak of the low-ionization lines (also coincident with the radio source). The peak of the \ion{He}{II} $\lambda$ 4686 emission corresponds to the location of the ULX. The dotted vertical lines correspond to the extent of the continuum emission from the star-forming complex. The magenta dashed vertical lines represent the 90\% confidence limits for the position of the ULX.}
  \label{vlt_spectra_x2_norm}
\end{figure}

\begin{figure}
\vspace{-0.7cm}
\hspace{-0.4cm}
\includegraphics[height=1.25\columnwidth, angle=270]{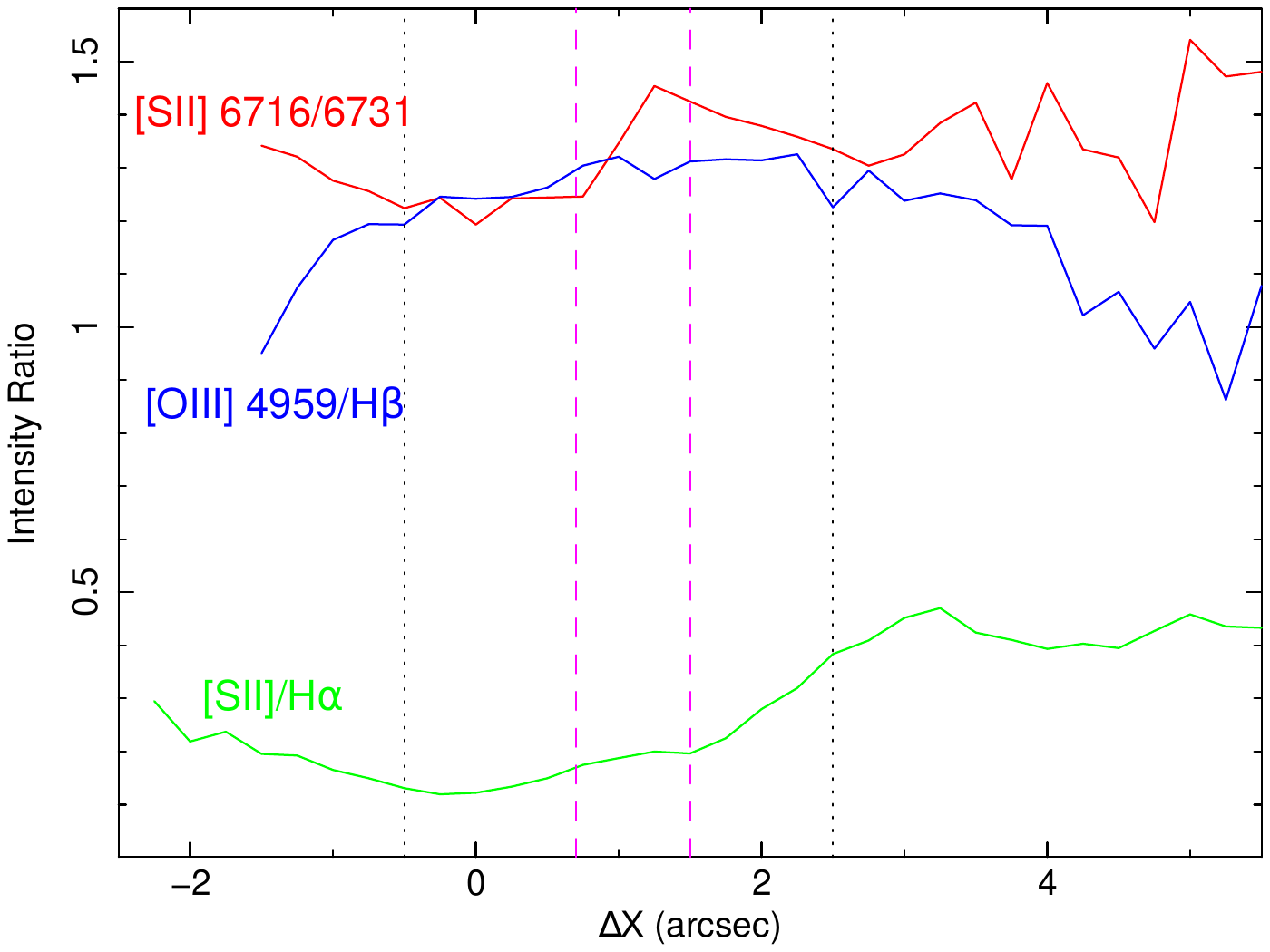}\\[-35pt]

\hspace{-0.4cm}
\includegraphics[height=1.25\columnwidth, angle=270]{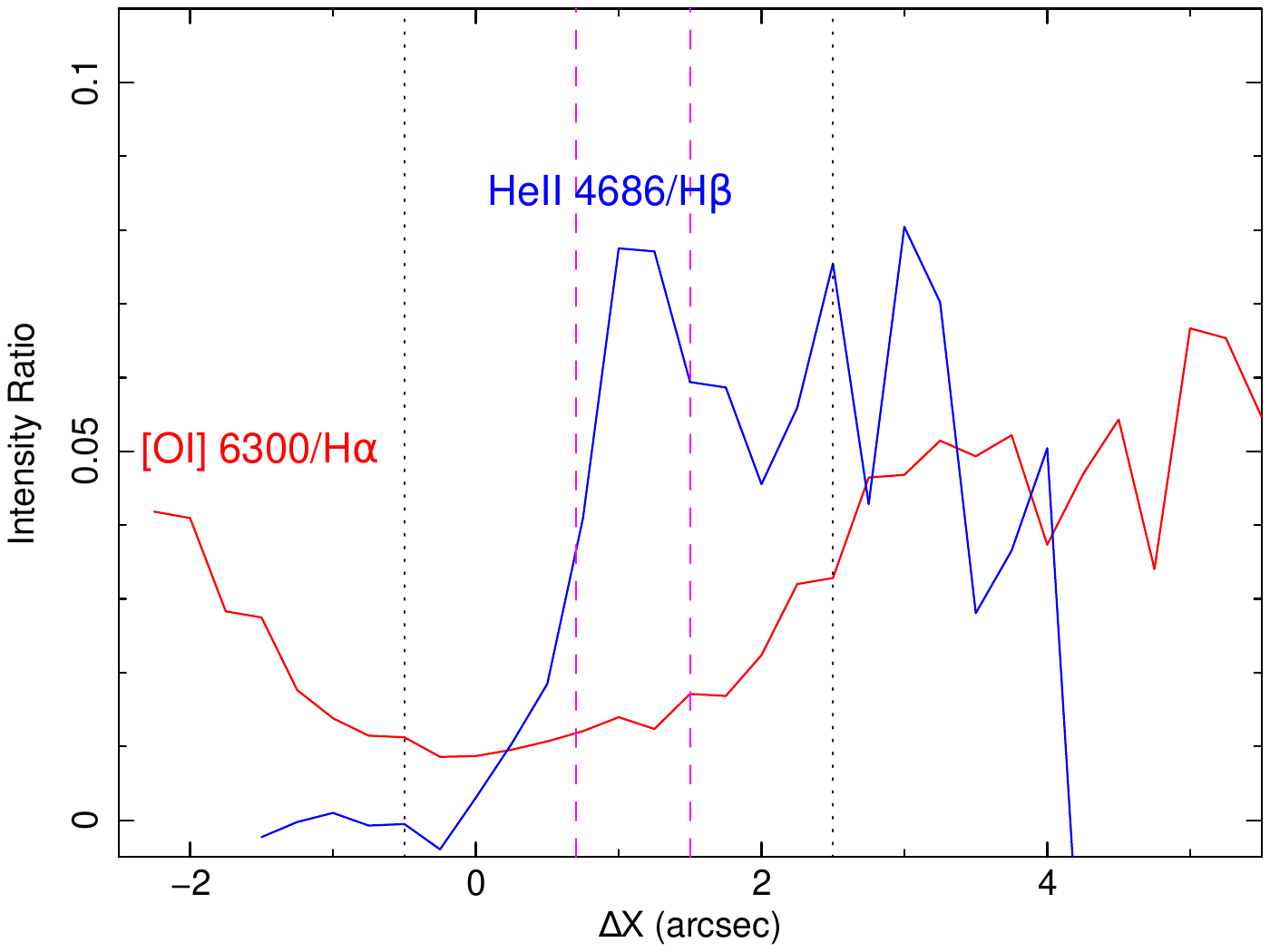}\\[-8pt]
 \caption{Both panels show a selection of line ratios of a few diagnostic lines, plotted along the north-south spatial direction. As in Figure 8, the $\Delta X= 0$ position on the X axis corresponds to the peak of the low-ionization lines. The magenta dashed vertical lines represent the 90\% confidence limits for the position of the ULX. The dotted vertical lines correspond to the extent of the continuum emission from the star-forming complex. The line-emitting nebula extends  a further $\approx$1$^{\prime\prime}$ south of the continuum-emitting region, and at least $\approx$3$^{\prime\prime}$ north of it. The nebular emission within the star-forming region (between the dotted lines) is consistent with an \ion{H}{2} region. The higher [\ion{S}{II}]/H$\alpha$ and [\ion{O}{I}]/H$\alpha$ ratios in the optically thin gas north of the ULX may be explained by the lower intensity of ionizing photons away from the southern star cluster.}
  \label{vlt_spectra_x2_ratios}
\end{figure}

\subsection{Total H$\alpha$ emission around X-2}
We used the FORS2 300V grism spectrum (average of the two 600-s exposures) to measure the observed flux on the slit. We found $F({\mathrm{H}}\alpha)_{\mathrm{sc,abs}}   = (12.3\pm0.1) \times 10^{-15}$ erg cm$^{-2}$ s$^{-1}$ on the 1$^{\prime\prime}$ slit around the southern star cluster, coincident with the radio detection. Because the seeing is approximately the same as the slit width, we need to take into account that a fraction of light from the cluster does not fall on the slit. 
We estimated this fraction by inspecting the acquisition images taken before and after the two 300V spectra, and determining the fraction of emission from bright, isolated stars that falls within a 1$^{\prime\prime}$-wide rectangular box. One of the two acquisition images was taken in slightly worse seeing conditions (1\farcs15) than the spectra, the other image had better seeing (0\farcs65); thus we took a weighted average of the two estimates. We also verified the result by blurring the {\it HST}/WFPC2 image in the F606W filter to simulate the ground-based seeing.
We obtain that $\approx$35\% of the light from isolated point-like sources is lost off the slit in the 300V spectra. However, in the case of the slit extraction around the southern star cluster, this loss is partly compensated by the additional emission from the star-forming complex east and west of the slit that ends up into the slit. Taking both effects into account, we estimate the total absorbed flux from the star cluster is $F({\mathrm{H}}\alpha)_{\mathrm{sc}}  \approx 15 \times 10^{-15}$ erg cm$^{-2}$ s$^{-1}$. 

Next, we need to correct for dust reddening. As mentioned earlier (Section 3.4), the Balmer decrement of 4.08 at the position of the star cluster suggests $E(B-V) \approx 0.36$ mag, that is $A_{{\mathrm H}\alpha} \approx 0.91$ mag, corresponding to $F({\mathrm{H}}\alpha)_{\mathrm{unabs}} \approx 2.31 F({\mathrm{H}}\alpha)_{\mathrm{abs}}$. Thus, our best estimate for the H$\alpha$ emission of the brightest cluster is $F({\mathrm{H}}\alpha)_{\mathrm{sc,unabs}}  \approx 3.6 \times 10^{-14}$ erg cm$^{-2}$ s$^{-1}$ and a luminosity $L({\mathrm{H}}\alpha)_{\mathrm{sc,unabs}}  \approx 5.0 \times 10^{38}$ erg s$^{-1}$. The 5.5-GHz flux density for the free-free radio emission associated with this Balmer flux is $F_{\rm 5.5GHz,ff} \approx 40 \mu$Jy \citep{caplan86}. The observed 5.5-GHz flux density (unresolved and spatially coincident with the star cluster) is $F_{\rm 5.5GHz,ff} = (138 \pm 15) \mu$Jy (Section 3.3). We conclude that free-free emission alone is a significant but not dominant component of the radio source in the X-2 complex.

Furthermore, we estimate the total H$\alpha$ luminosity from the whole star-forming clump, defined as the $\approx$2$^{\prime\prime} \times 3^{\prime\prime} (\approx 100 {\mathrm{~pc}} \times 150$ pc) region with both starlight continuum and nebular line emission. The FORS2 spectra show a flux $F({\mathrm{H}}\alpha)_{\mathrm{tot,abs}}   = (22\pm1) \times 10^{-15}$ erg cm$^{-2}$ s$^{-1}$ on the slit. From the VLT/FORS2 acquisition image taken the same night and with the same seeing, we estimate that the emission along the slit is $\approx$60\% of the total emission from the star-forming clump. Thus, we estimate that $F({\mathrm{H}}\alpha)_{\mathrm{tot,abs}}   = (3.7\pm0.5) \times 10^{-14}$ erg cm$^{-2}$ s$^{-1}$. The emission-weighted average Balmer decrement along the slit is $\approx$3.9 (running from $\approx$4.1 at the younger, brighter southern end to $\approx$3.5 at the slightly older, fainter northern end). This corresponds to $E(B-V) \approx 0.31$ mag, $A_{{\mathrm H}\alpha} \approx 0.79$ mag, and $F({\mathrm{H}}\alpha)_{\mathrm{unabs}} \approx 2.06 F({\mathrm{H}}\alpha)_{\mathrm{abs}}$. Thus, our best estimate for the dereddened H$\alpha$ flux of the star-forming region is $F({\mathrm{H}}\alpha)_{\mathrm{tot,unabs}}  \approx (7.6\pm 1.0) \times 10^{-14}$ erg cm$^{-2}$ s$^{-1}$ and a luminosity $L({\mathrm{H}}\alpha)_{\mathrm{tot,unabs}}  \approx (1.1\pm0.2) \times 10^{39}$ erg s$^{-1}$.


\begin{table}
\caption{Comparison of \ion{He}{II} $\lambda 4686$ and X-ray luminosities in a sample of nearby ULX associated with proven or candidate photo-ionized nebulae.} 
\begin{center}  
\begin{tabular}{lcccc} 
 \hline 
\hline \\[-8pt]
  ULX   &   $d$  &   $L_{4686}^{a}$  & $L_{0.3-10}^{b}$ & 
  Ratio \\
  &   (Mpc)   & $\left(10^{36} {\rm erg~s}^{-1}\right)$     
  & $\left(10^{39} {\rm erg~s}^{-1}\right)$   &  $\left(10^{-4}\right)$  \\ %
    \hline\\[-7pt]
 NGC\,7424 X-2  &  10.8 &  $\sim$8$^c$ & $\sim$7 (0.6--8)  & $\sim$11  \\
 Ho II X-1  & 3.05  &  2.7 & $\sim$7 (2--18) & $\sim$4  \\
 M\,81 X-6$^d$  & 3.6  & $\gtrsim$0.11$^e$  & $\sim$5 (2--9)  &  $\gtrsim$0.2 \\
 NGC\,5408 X-1  & 4.8  & 1.1  & $\sim$7 (2--12)  &  $\sim$2 \\
 NGC\,6946 X-1$^f$  & 7.7  &  40 &  $\sim$9(6--20) &  $\sim$45$^g$ \\
 M\,51 X107$^h$ & 8.6  & 0.8  & 2  &  4 \\
 1\,Zw\,18 X-1 &  19 & 123  & 3.2  &  380$^i$ \\
 Mrk 1434 X-N$^j$ &  31 & 86  & 16 &  54$^k$ \\
   \hline
\vspace{-0.5cm}
\end{tabular}
\end{center}
\begin{flushleft} 
$^a$: Extinction-corrected luminosity of the nebular He II $\lambda 4686$ line. Line-of-sight Galactic extinction is assumed, when an intrinsic value was not given in the literature;\\
$^b$: unabsorbed (0.3--10)-keV luminosity of the ULX. As ULXs are typically variable, a ''characteristic'' value (not a precise mathematical average) is reported, as well as a range of luminosities found at different epochs;\\
$^c$: assuming that the He {\footnotesize {II}} $\lambda 4686$ photons see an average extinction $A_V \approx 0.67$ mag (based on the Balmer decrement);\\
$^d$: CXO\,J095532.9$+$690033;\\
$^e$: includes only the emission integrated along the slit;\\
$^f$: the nebula is best known as NGC\,6946 MF16;\\
$^g$: He {\footnotesize {II}} $\lambda 4686$ emission requires both a photo-ionized and a shock-ionized component;\\
$^h$: CXOM51\,J132940.0$+$471237\\
$^i$: He {\footnotesize {II}} $\lambda 4686$ emission probably not powered (only) by X-1; ionization by a hot superbubble proposed by \citep{oskinova22} but ruled out by \cite{franeck22};\\
$^j$: CXO\,J103410.1$+$580349;\\
$^k$: He {\footnotesize {II}} $\lambda 4686$ emission probably not powered (only) by X-N;\\
References: for NGC\,7424 X-2: this work; 
for $L_{4686}$ from Ho II X-1: \cite{moon11,lehmann05,kaaret04,pakull02}; 
for $L_{0.3-10}$ from Ho II X-1: \cite{barra23,gurpide21,walton15,sutton13,grise10}; 
for $L_{4686}$ from M\,81 X-6: \cite{moon11}
for $L_{0.3-10}$ from M\,81 X-6: \cite{bernadich22,evans20,webb20,swartz03};
for $L_{4686}$ from NGC\,5408 X-1: \cite{kaaret09};
for $L_{0.3-10}$ from NGC\,5408 X-1: \cite{garcia13,sutton13,grise13,kaaret09};
for $L_{4686}$ from NGC\,6946 X-1: \cite{abolmasov08};
for $L_{0.3-10}$ from NGC\,6946 X-1: \cite{earnshaw19,middleton15,sutton13,roberts03}; see also \cite{kaaret10} for the ionizing UV flux in this source;
for $L_{4686}$ from M\,51 X107: \cite{urquhart18}; 
for $L_{0.3-10}$ from  M\,51 X107: \cite{urquhart16,kuntz16}; 
for $L_{4686}$ from 1\,Zw\,18: \cite{rickards21,kehrig21,kehrig15}; 
for $L_{0.3-10}$ from 1\,Zw\,18 X-1: \cite{thuan04}; 
for $L_{4686}$ from Mrk 1434: \cite{thygesen23}; 
for $L_{0.3-10}$ from Mrk 1434 X-N: \cite{thygesen23,lemons15}.
\\
\end{flushleft}
\end{table}

\begin{figure}
\vspace{-0.7cm}
\hspace{-0.4cm}
\includegraphics[height=1.25\columnwidth, angle=270]{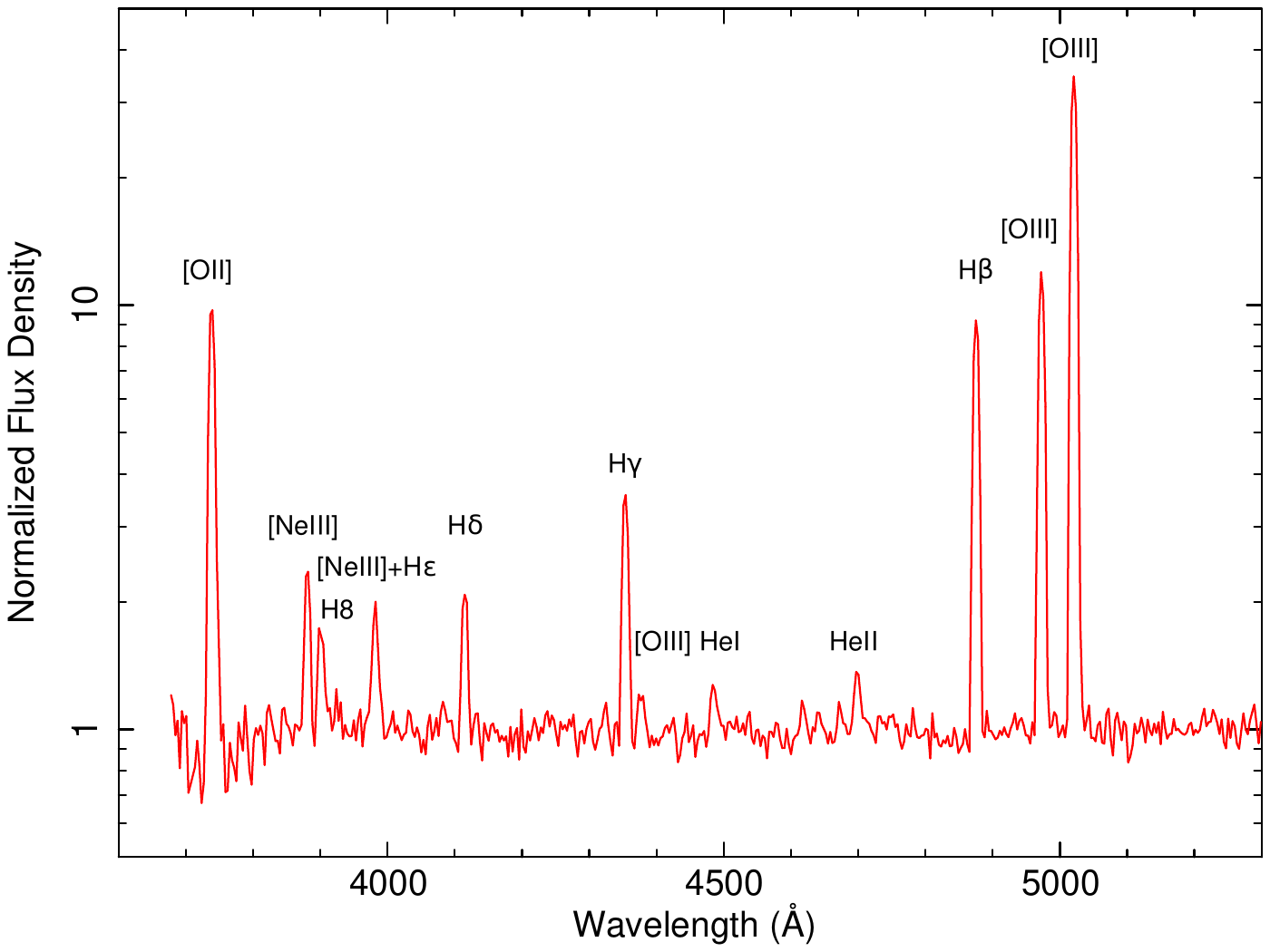}\\[-35pt]

\hspace{-0.4cm}
\includegraphics[height=1.25\columnwidth, angle=270]{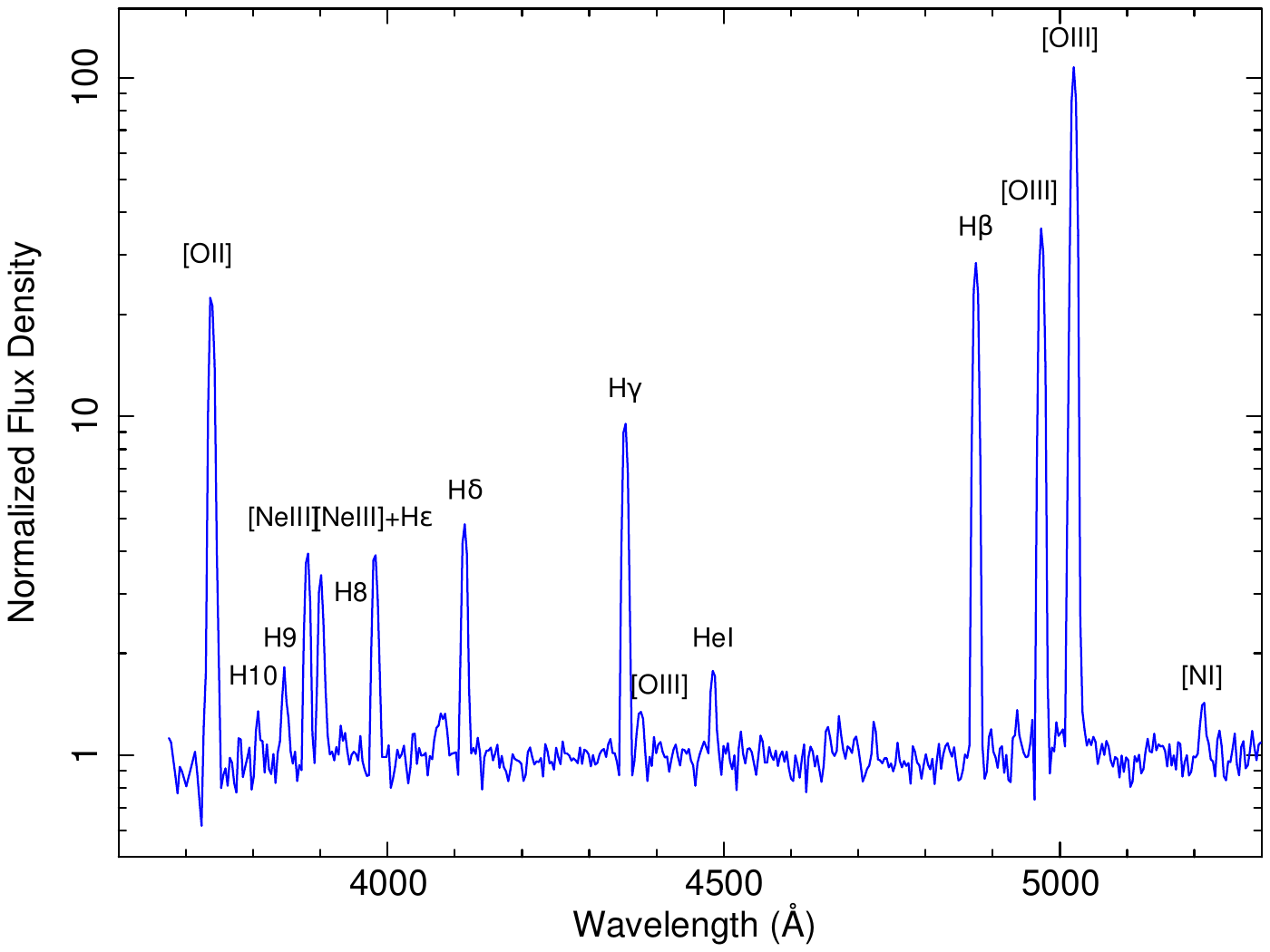}\\[-35pt]

\hspace{-0.4cm}
\includegraphics[height=1.25\columnwidth, angle=270]{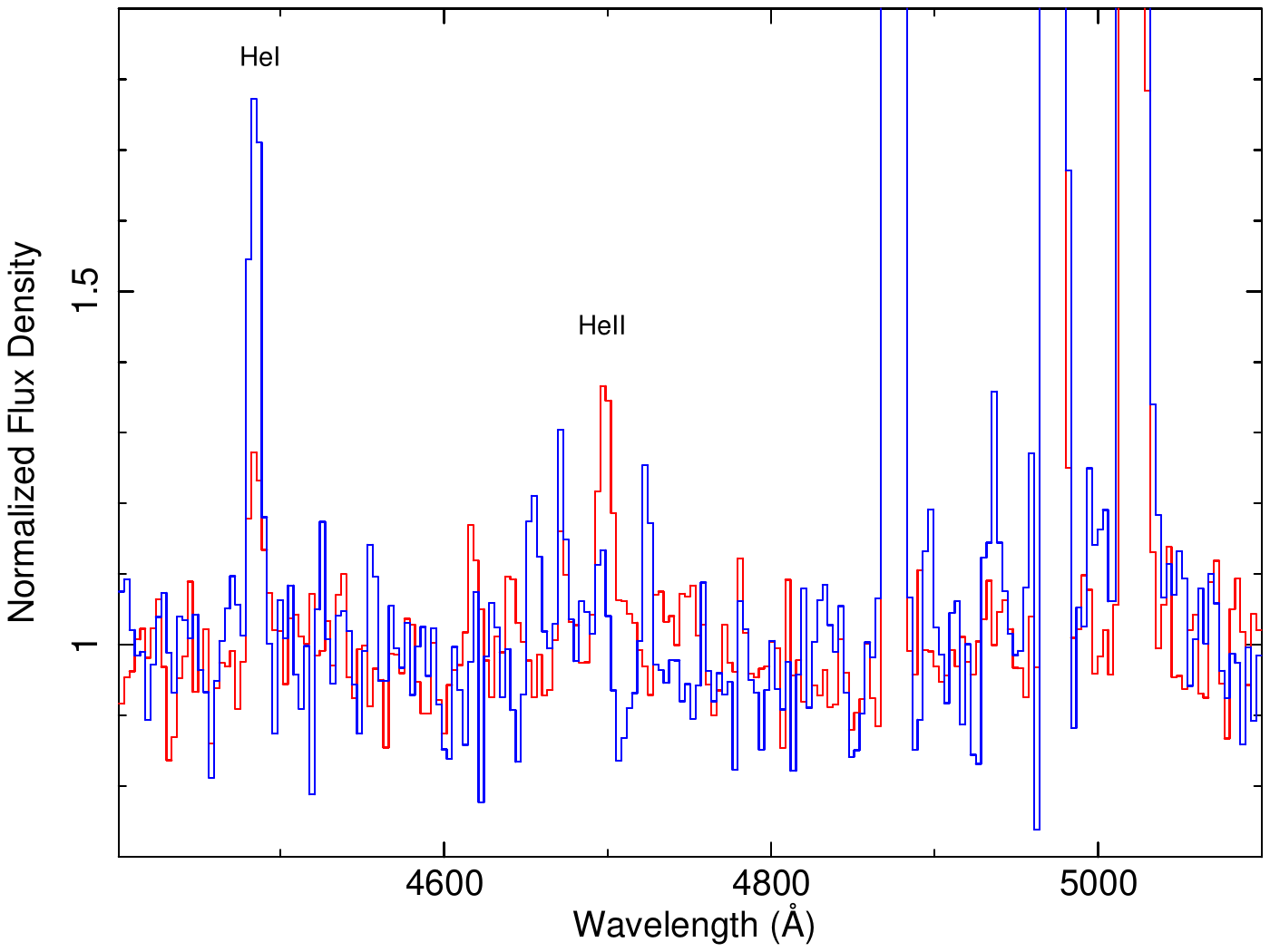}\\[-8pt]
\caption{Top panel: blue section of the VLT/FORS2 300V spectrum of the northern part of the star-forming complex, with a spatial extent of $\Delta Y = 0\farcs5$ around the {\it Chandra} position of X-2. The spectral dispersion is 3.3 \AA\ per pixel. The instrumental resolution is 9.4 \AA\ FWHM.
Middle panel: the same section of the VLT/FORS2 300V spectrum, for the southern part of the star-forming complex, with a spatial extent of $\Delta Y = 0\farcs5$ around the peak of the low-ionization lines.
Bottom panel: zoomed-in view of the two spectra shown in the top and middle panel, to highlight the stronger contribution of \ion{He}{II} compared with \ion{He}{I} around the X-ray location (red histogram), and the absence of \ion{He}{II} in the bottom part of the nebula (blue histogram). 
}
  \label{fors2_1d_blue}
\end{figure}

\begin{figure}
\vspace{-0.7cm}
\hspace{-0.4cm}
\includegraphics[height=1.25\columnwidth, angle=270]{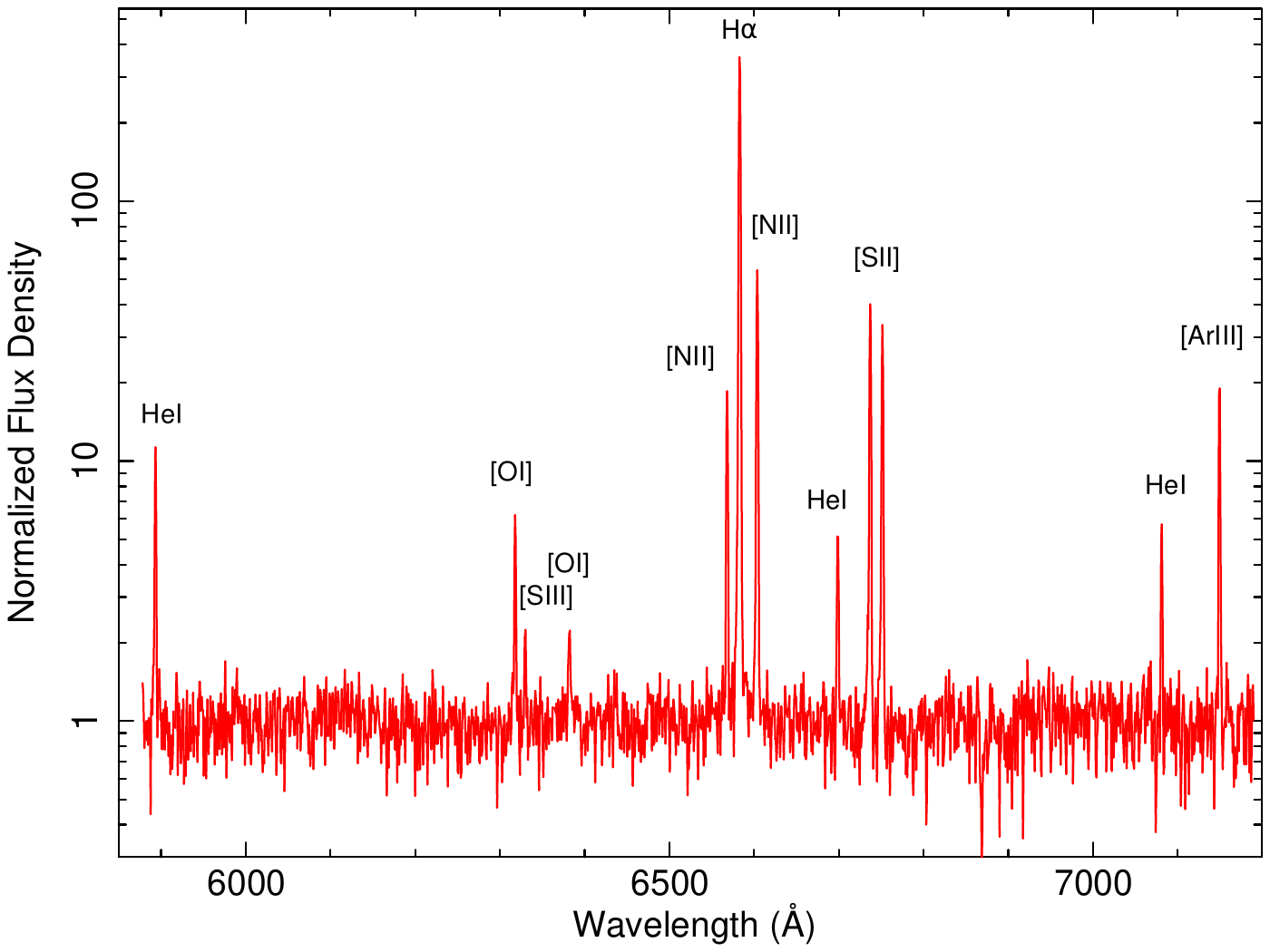}\\[-35pt]

\hspace{-0.4cm}
\includegraphics[height=1.25\columnwidth, angle=270]{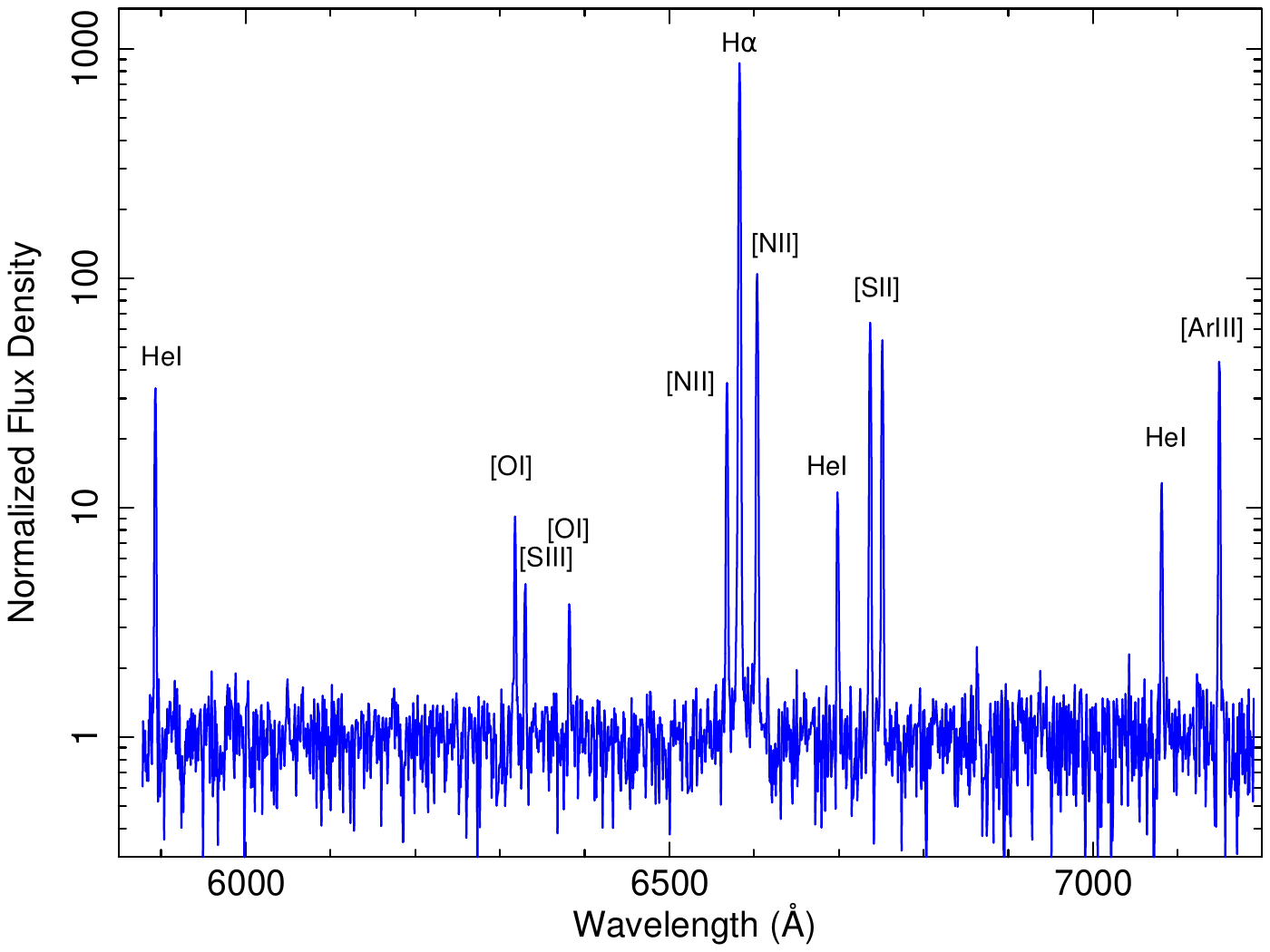}\\[-35pt]

\hspace{-0.4cm}
\includegraphics[height=1.25\columnwidth, angle=270]{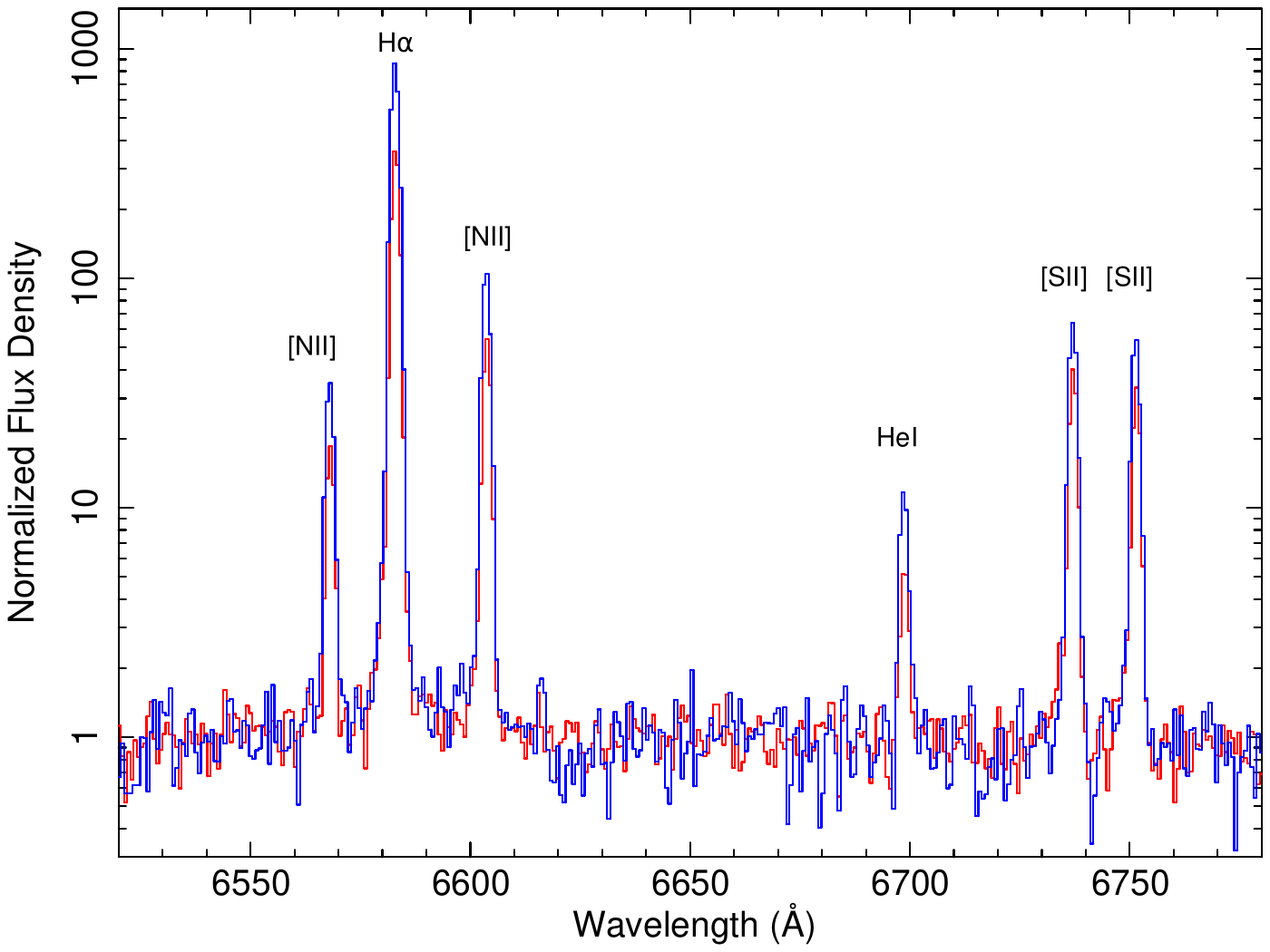}\\[-8pt]
\caption{Top panel: VLT/FORS2 1200R spectrum of the northern part of the star-forming complex, with a spatial extent of $\Delta Y = 0\farcs5$ around the {\it Chandra} position of X-2. The spectral dispersion is 0.74 \AA\ per pixel.
Middle panel: VLT/FORS2 1200R spectrum, for the southern part of the star-forming complex, with a spatial extent of $\Delta Y = 0\farcs5$ around the peak of the low-ionization lines.
Bottom panel: zoomed-in view of the two spectra shown in the top and middle panel (red and blue histograms, respectively).
}
  \label{fors2_1d_red}
\end{figure}

\subsection{He II emission near X-2 and other line diagnostics}

Our VLT spectra reveal for the first time \ion{He}{II} $\lambda$4686 emission peaking around the {\it Chandra} position for X-2 (Figure 8, Table 4). This is strongly indicative of X-ray photo-ionization near the ULX, while the rest of the star-forming complex is consistent with a normal \ion{H}{II} region (UV photo-ionized by OB stars). The He$^{++}$ emission extends $\approx$3$^{\prime\prime} \approx 150$ pc to the north of the ULX (Figures 8,9), that is $\approx$1\farcs5 into the purely nebular region, but is much weaker or undetected to the south of the ULX, towards the peak Balmer emission and radio source. A comparison between the two spectra extracted around the position of the radio source and around the position of the ULX show (Figure 10, Table 4) that \ion{He}{I} $\lambda$4471 is stronger than \ion{He}{II} $\lambda$4686 in the former, and weaker in the latter.

Apart from the He lines, other emission lines do not show any dramatic difference at the ULX location compared with the southern cluster (Figures 10,11). Low-ionization line ratios [S {\footnotesize{II}}] $\lambda$6716+6731/H$\alpha$ and [O {\footnotesize{I}}] $\lambda$6300/H$\alpha$ are low (Figure 9), consistent with a normal H {\footnotesize II} region, across all the nebula. Both ratios have a minimum at the position of the southern star cluster and increase away from the main source of ionization.  We suggest that this is due to a larger fraction of O and S being more ionized at the denser, brighter southern end of the star-forming clump. An inverse correlation of [S {\footnotesize{II}}] $\lambda$6716+6731/H$\alpha$ with H$\alpha$ intensity and electron density is seen also in the Galactic disk ({\it e.g.}, \citealt{hill14}).

The density diagnostic ratio [S {\footnotesize{II}}] 6716/6731 $\approx 1.35 \pm 0.05$ near the position of the ULX (Figure 9, Table 5) is consistent with moderate or low density $n_e \lesssim $ a few $\times 10$ cm$^{-3}$  \citep{osterbrock06}. The lowest value (ratio $\approx$1.4) is exactly at the ULX position, a possible indication of a lower-density or higher ionization cavity around the X-ray source. A marginally significant density increase is found instead at the southern location, where [S {\footnotesize{II}}] 6716/6731 $\approx 1.23 \pm 0.05$, consistent with $n_e \approx 100$ cm$^{-3}$. If the steep-spectrum radio source is a young SNR, the enhanced ISM density in that region may explain its high luminosity.

The temperature diagnostic ratio [O {\footnotesize{III}}] (4959+5007)/4363 is $\approx$($150 \pm 20$) in the southern part of the clump, and $\approx$($100 \pm 20$) around the ULX. This corresponds to an electron temperature range $\approx$10,000--13,000 K between the two regions. The low signal-to-noise ratio of the [O {\footnotesize{III}}] $\lambda$4363 line prevents a more accurate spatially resolved study.

\begin{table*}
\caption{Best-fitting parameters of the {\it Chandra}/ACIS-S spectra of X-1, fitted with the Cash statistics. Uncertainties for one interesting parameter are reported at the confidence interval of $\Delta C = \pm$2.70: this is asymptotically equivalent to the 90\% confidence interval in the $\chi^2$ statistics.
The Galactic absorption is fixed at $N_{\rm {H,Gal}} = 8.6 \times 10^{19}$ cm$^{-2}$.} 
\vspace{-0.3cm}
\begin{center}  
\begin{tabular}{lccc} 
 \hline 
\hline \\[-8pt]
  Model Parameters      &      \multicolumn{3}{c}{Values} \\
  & 2002 May 21       &       2002 June 11   &   2020 December 02  \\ %
\hline\\[-9pt]
   \hline\\[-9pt]
\multicolumn{4}{c}{{\it tbabs} $\times$ {\it tbabs} $\times$ {\it powerlaw}}\\
\hline\\[-5pt]
   $N_{\rm {H,int}}$   ($10^{22}$ cm$^{-2}$)   &  $ 0.09^{+0.03}_{-0.03} $ &     $ 0.13^{+0.04}_{-0.04}$ & $ 0.00^{+0.34}_{-0.00}$\\[4pt]
   $\Gamma$   & $ 1.86^{+0.15}_{-0.15} $ & $ 1.58^{+0.12}_{-0.12} $ & $ 1.70^{+0.46}_{-0.31} $ \\[4pt]
   $N_{\rm {po}}^a$  & $ 6.13^{+0.81}_{-0.70} $ & 
   $ 8.71^{+1.04}_{-0.91} $ & $ 5.13^{+3.33}_{-1.31}$\\[4pt]
   C-stat/dof     &      $219.4/237$ (0.93)            &       $253.6/303$ (0.84) & $52.1/70$ (0.74) \\[4pt]
   $f_{0.3-10}$ ($10^{-13}$ erg cm$^{-2}$ s$^{-1}$)$^b$ & $ 3.16^{+1.09}_{-1.09} $ & $ 5.89^{+1.07}_{-1.07}$ & $ 3.55^{+1.17}_{-1.17}$ \\[4pt]
   $L_{0.3-10}$ ($10^{39}$ erg s$^{-1}$)$^c$ &  $ 5.30^{+0.38}_{-0.35}$ & $ 9.43^{+0.68}_{-0.63}$ & $ 4.95^{+1.00}_{-0.83} $   \\[4pt]
\hline 
\vspace{-0.5cm}
\end{tabular}
\end{center}
\begin{flushleft} 
$^a$: units of $10^{-5}$ photons keV$^{-1}$ cm$^{-2}$ s$^{-1}$ at 1 keV.\\
$^b$: observed fluxes in the 0.3--10 keV band\\
$^c$: isotropic unabsorbed luminosities in the 0.3--10 keV band, defined as $4\pi d^2$ times the de-absorbed fluxes.\\
\end{flushleft}
\end{table*}

\begin{figure}
\vspace{-1.0cm}
\hspace{-0.4cm}
\includegraphics[height=1.25\columnwidth, angle=270]{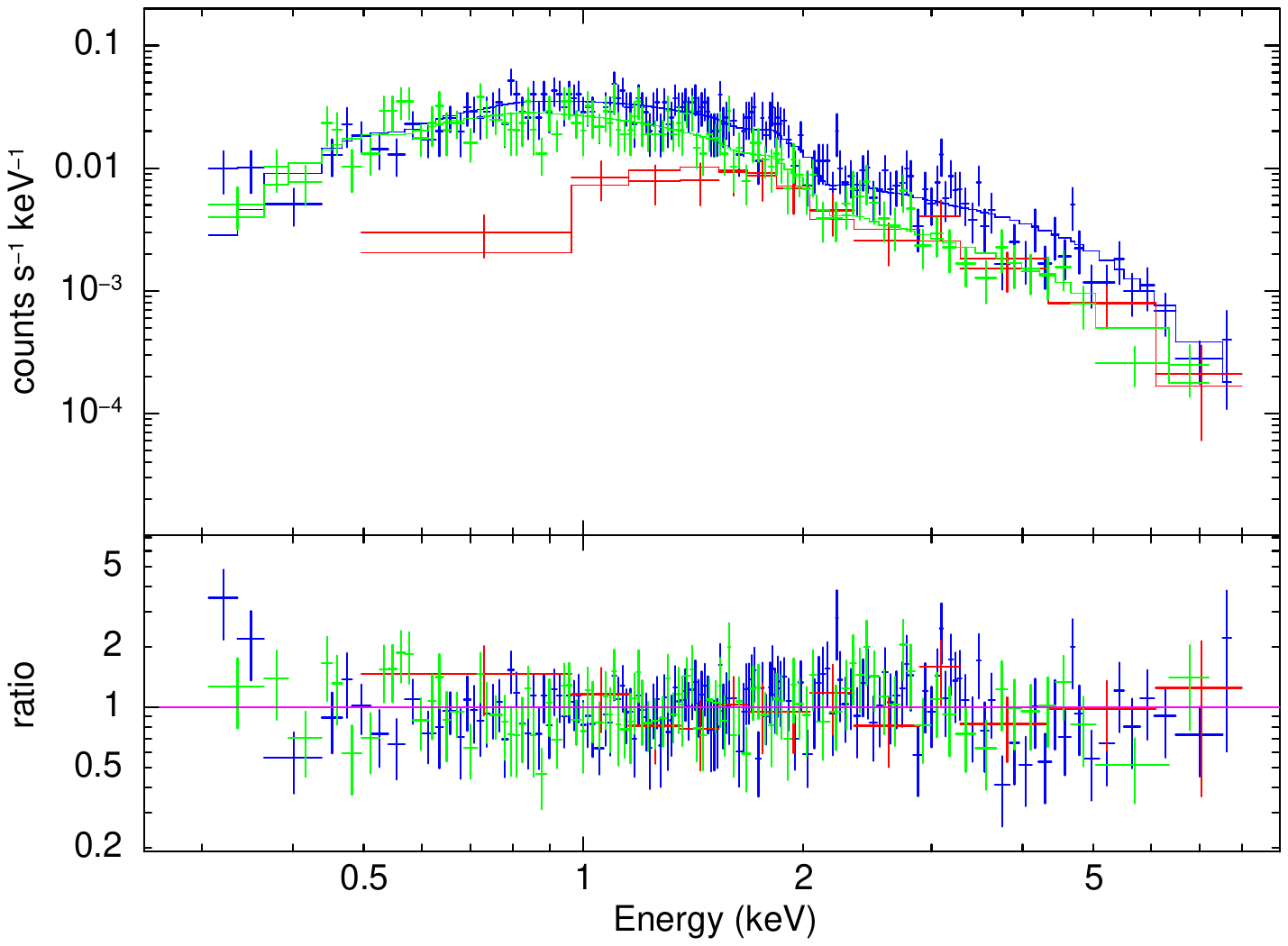}\\[-3pt]
\caption{Datapoints and data/model ratios for the {\it Chandra}/ACIS spectrum of X-1. Green is for ObsID 3495, blue for ObsID 3496 and red for ObsID 23572. The model is an absorbed powerlaw (Table 3). The spectra were fitted with the Cash statistics. The datapoints were rebinned to a minimum signal to noise ratio of 2.7 for plotting purposes only.
}
  \label{x1_chandra}
\end{figure}

\subsection{Dereddened luminosity of the He$^{++}$ nebula}

By analogy with the H$\alpha$ emission discussed in Section 3.5, we estimate now the extinction-corrected luminosity of the He {\footnotesize {II}} $\lambda$4686 emission, to place it in the context of other He$^{++}$ ULX nebulae in the literature. The integrated flux along the slit is $F(4686)_{\mathrm{tot,abs}} = (1.7\pm0.1) \times 10^{-16}$ erg cm$^{-2}$ s$^{-1}$. We then need to estimate the fraction of emission lost outside the slit. As discussed before, for a point-like emission source centred in the middle of the slit, we would lose $\approx$35\% of the flux. This gives us a total flux $F(4686)_{\mathrm{tot,abs}} = (2.6\pm0.2) \times 10^{-16}$ erg cm$^{-2}$ s$^{-1}$. However, this is a lower limit, because we know that the He$^{++}$ region extends $\approx$3$^{\prime\prime}$ in the north-south direction along the slit (Figure 8); thus, it is plausible that a tail of emission extends also a similar amount in the east-west direction. Using analogous arguments to those applied to H$\alpha$ ({\it i.e.}, $\approx$40 per cent loss), we find a more plausible estimate of the total He {\footnotesize {II}} $\lambda$4686 emission as $F(4686)_{\mathrm{tot,abs}} = (2.8\pm0.2) \times 10^{-16}$ erg cm$^{-2}$ s$^{-1}$. As a further check, we analyzed a spectrum of a well-calibrated Wolf-Rayet star (WR9 in the Small Magellanic Cloud: \citealt{crowther06,massey03}) taken with the FORS2 300V grism at the end of the same night (2001 September 01), and used it as a flux calibrator for the  He {\footnotesize {II}} $\lambda$4686 emission. From that, we confirm an observed flux  $F(4686)_{\mathrm{tot,abs}} \sim 3 \times 10^{-16}$ erg cm$^{-2}$ s$^{-1}$.

Then, we need to correct for extinction.  The H$\alpha$/H$\beta$ Balmer ratio around the ULX position is $\approx$3.54 (Table 4): this implies $E(B-V) \approx 0.22$ mag, $A_{4686} \approx 0.82$ mag, and $F(4686)_{\mathrm{unabs}} \approx 2.12 F(4686)_{\mathrm{abs}}$. Hence, we obtain a dereddened He {\footnotesize {II}} $\lambda$4686 flux $F(4686)_{\mathrm{tot,unabs}}  \approx (5.9\pm0.5) \times 10^{-16}$ erg cm$^{-2}$ s$^{-1}$ and a best-estimate luminosity $L(4686)_{\mathrm{sc,unabs}}  \approx (8.3\pm0.7) \times 10^{36}$ erg s$^{-1}$. 

The luminosity of the He$^{++}$ nebula is consistent with those seen in other ULXs with roughly similar X-ray luminosities, with $L_{\rm X}/L_{4686} \sim $ a few $10^{-4}$ (Table 4). In fact, it is a factor of 3 more luminous and more extended than the prototypical He$^{++}$ nebula around the ULX in the Holmberg II galaxy, and several times more luminous that the X-ray photo-ionized nebula around NGC\,5408 X-1.

\begin{figure}
\centering
\includegraphics[width=\columnwidth]{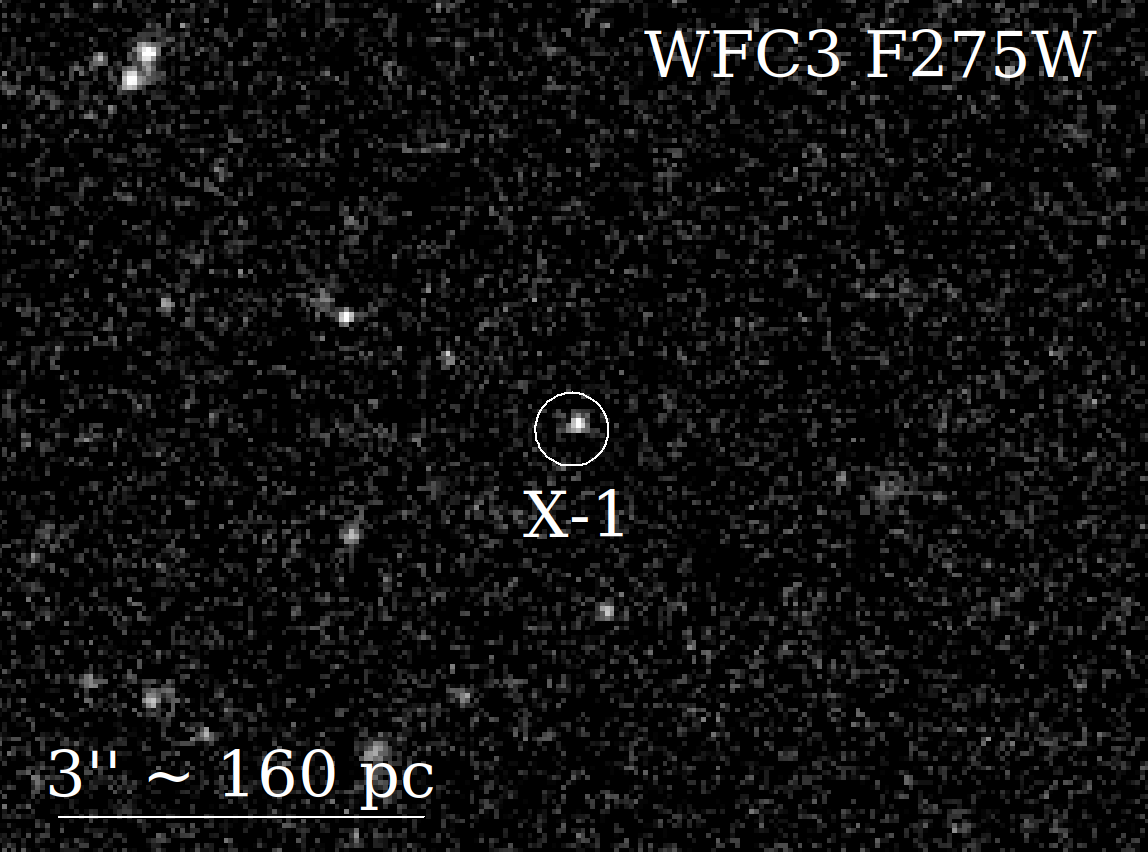}
\caption{UV counterpart of the X-1 ULX ({\it HST}/WFC3 F275W image). There is only one blue star inside the 90\% error radius of the X-ray position (radius of 0\farcs3). Its brightness and colour are consistent with a B8 supergiant with a mass $\sim$9 $M_\odot$ at an age of $\sim$30 Myr. 
}
\label{x1_wfc3}
\end{figure}

\begin{figure}
\centering
\includegraphics[width=\columnwidth]{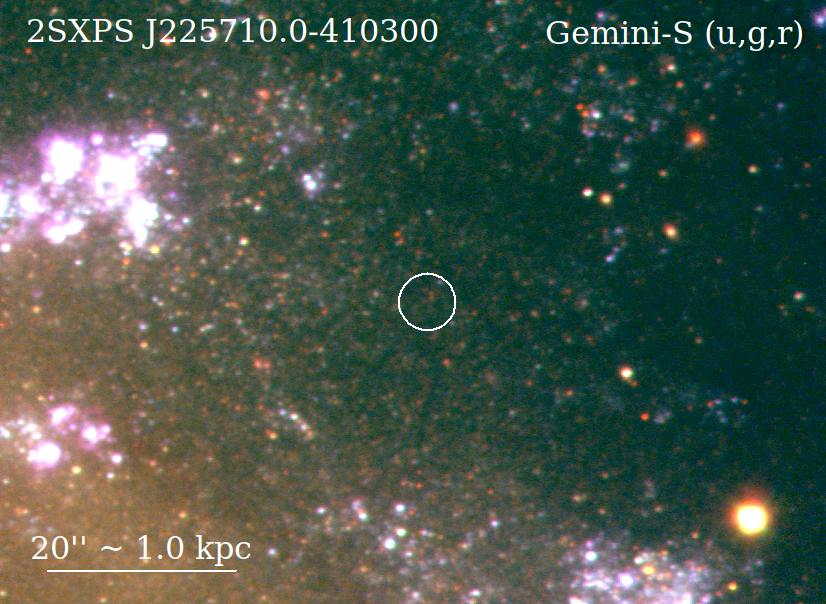}
\caption{Gemini GMOS field around the transient ULX 2SXPS\,J225710.0$-$410300 ({\it Swift}/XRT error circle overplotted in white, with a radius of 3$^{\prime\prime}$). 
}
\label{swift_transient}
\end{figure}

\subsection{Multiband properties of the X-1 ULX}

Although the main focus of our study was the X-2 ULX and its environment, we also provide an update on two other ULXs in this galaxy.
X-1 \citep{soria06} remained in an ultraluminous regime in the 2020 {\it Chandra} observation, at $L_{\rm X} \approx 5 \times 10^{39}$ erg s$^{-1}$ (Table 5), similar to the value measured in 2002 May but a a factor of two lower than in 2002 June. No significant deviation from a simple power law (Figure 12) was recorded in any of the three spectra.

The field around X-1 is included in two deep {\it HST}/WFC3-UVIS observations from 2016 (Table 1), but not in the earlier WFPC2 observations.
There is only one point-like optical source inside the {\it Chandra} error circle of the ULX position (Figure 13). 
We measured apparent brightness $m_{275W} = (24.65 \pm 0.05)$ mag and $m_{336W} = (24.69 \pm 0.05)$ mag (both values are in the Vegamag system). We then corrected the observed values for a line-of-sight Galactic extinction (taken from NED)
of $A_{275W} \approx 0.06$ mag and $A_{336W} \approx 0.05$ mag, and applied a distance modulus of 30.17 mag (Section 1). We obtained absolute magnitudes $M_{275W} = (-5.58 \pm 0.05)$ mag and $M_{336W} = (-5.53 \pm 0.05)$ mag.

We used the Padova isochrones \citep{bressan12} 
to estimate the properties of a single solar-metallicity star with such colours. It is consistent (approximately 90 per cent confidence level) with a blue supergiant of age $(28 \pm 3)$ Myr, a mass of $9.2^{+0.6}_{-0.4} M_{\odot}$, a temperature of about $(11,100 \pm 1400)$ K and a radius of about $(30 \pm 10) R_{\odot}$. However, there is an ongoing debate \citep[{\it e.g.},][]{lian11,gladstone13} on whether blue optical counterparts of ULXs are their donor stars, or the emission from the irradiated outer disk (which for plausible orbital parameters has similar size and temperature as a B supergiant), or a mix of the two components. From only two near-UV bands, we cannot discriminate between the two possibilities. We cannot also definitively rule out the possibility that the point source is a background quasar. However, considering the rarity of background sources at the observed X-ray flux ($\sim$1 degree$^{-2}$: \citealt{luo17,cappelluti09}) and the high X-ray/optical flux ratio of $\sim$500--1000, we consider this scenario very unlikely. Finally, we inspected the ATCA data but found no radio emission from X-1 or its surroundings, to a 4-$\sigma$ upper limit of $\approx$50 $\mu$Jy at 5.5 GHz.


\subsection{Another transient ULX found with {\it Swift}}
The {\it Niels Gehrels Swift Observatory} observed NGC\,7424 twice, with snapshot observations: on 2008 July 23 (698-s exposure time for the X-Ray Telescope, XRT) and on 2008 November 25 (2093 s). We retrieved the data from NASA's Heasarc public archive, and estimated count rates and fluxes of the detected sources using standard data analysis tools for {\it Swift}/XRT available online\footnote{\url{https://www.swift.ac.uk/user_objects/}}. We also compared the results of our re-analysis with those reported in the 2SXPS Catalogue of {\it Swift X-ray Telescope} Point Sources \citep{evans20} and found them consistent. The most important result of the {\it Swift}/XRT data is the discovery of a transient ULX $\approx$2$^{\prime}$ north-west of the nucleus, at R.A.(J2000) $= 22^h\,57^m\,10^{s}.02$, Dec.(J2000) $= -41^{\circ}\,03^{\prime}\,00\farcs5$ (90\% confidence radius $\approx$3$^{\prime\prime}$). The X-ray source (2SXPS\,J225710.0$-$410300, \citealt{evans20}) was seen in the 2008 November 25 observation with $\approx$19 net counts in the 0.3--10 keV band, but was not detected on 2008 July 23. Given the small number of counts, it is impossible to do any spectral analysis; however, with simple assumptions of a power-law spectrum, and column density $\sim$10$^{21}$ cm$^{-2}$, we estimate a luminosity $L_{\rm X} = (9 \pm 2) \times 10^{39}$ erg s$^{-1}$ on  2008 November 25, and $L_{\rm X} \lesssim 3 \times 10^{39}$ erg s$^{-1}$ on 2008 July 23 (using the 90\% confidence limits of \citealt{kraft91}). 2SXPS\,J225710.0$-$410300 was not detected in any of the three {\it Chandra} observations; we estimate an upper limit to its luminosity in 2002 of $L_{\rm X} \lesssim 10^{37}$ erg s$^{-1}$, from a stack of the ObsID 3495 and 3496 datasets.  
X-1 and X-2 are visible in both {\it Swift}/XRT observations, with unabsorbed luminosities (averaged over the two XRT datasets) of $L_{\rm X} = (7 \pm 2) \times 10^{39}$ erg s$^{-1}$ and $L_{\rm X} = (3 \pm 1) \times 10^{39}$ erg s$^{-1}$, respectively.

2SXPS\,J225710.0$-$410300 is outside the field of view of all {\it HST} observations of NGC\,7424. However, it is in the field of view of the Gemini GMOS images. There are two faint, red stars near the centre of the XRT error circle, both of them with apparent brightness $I \sim 24.5$ mag (Figure \ref{swift_transient}). The two stars are detected as a single star (DES\,J225709.98$-$410259.7) in the Dark Energy Survey Data Reslease 2 catalogue \citep{abbott21}. There is no radio detection in the ATCA data, down to a 4-$\sigma$ upper limit of $\approx$30 $\mu$Jy at 5.5 GHz. 

\subsection{eROSITA detection of X-1 but not X-2}
The field of NGC\,7424 was observed by the eROSITA telescope array on the {\it Spektrum Roentgen Gamma} ({\it SRG}) satellite \citep{predehl21}, on 2020 May 15 (MJD 58984), for a net (vignetted-corrected) exposure time of 101 s. The recently released eRASS1 catalogue \citep{merloni24} and associated sky maps show only one detected source in the galaxy, catalogued as 1eRASS J225728.6$-$410215. This source is clearly identifiable as X-1, within the position uncertainty (1$\sigma$ error of $\approx$2\farcs5 in R.A. and $\approx$3\farcs0 in Dec.). It has $\approx$15$\pm$4 net counts (detection likelihood of $\approx$41), corresponding to an absorbed 0.2--2.3 keV flux of $(1.4 \pm 0.4) \times 10^{-13}$ erg cm$^{-2}$ s$^{-1}$. This is the same flux observed in the {\it Chandra} observations of 2002 May and 2020 December, in the same band. Assuming a similar power-law model as in Table 5, we conclude that X-1 was at a 0.3--10 keV luminosity of $\sim$5 $\times 10^{39}$ erg s$^{-1}$ during the eROSITA observation. By contrast, X-2 was not detected by eROSITA. From an inspection of the sky map, we estimate that its 0.2--2.3 keV flux in 2020 May must have been at least 4 times lower than observed in 2020 December, and we place a rough upper limit of $\sim$2 $\times 10^{39}$ erg s$^{-1}$ to its 0.3--10 keV luminosity.


\begin{table*}
\caption{Selected properties of the three main star-forming regions in the spiral arms of NGC\,7424, identified from {\it WISE} images. The three regions have been labelled as in Figure 15, for simplicity. For each of the three young stellar complexes, we list its mid-IR brightness in the four {\it WISE} channels (Vegamag units),  its brightest X-ray source (with its indicative X-ray luminosity from {\it Chandra}), its brightest radio source (with 5.5 GHz and 9.0 GHz flux densities from the ATCA), and its star-formation rate (from {\it WISE} W3 and W4).} 
\begin{center}  
\begin{tabular}{lcccccccccc} 
 \hline 
\hline \\[-8pt]
  Region$^a$  & W1 & \hspace{-0.2cm}W2 & \hspace{-0.2cm}W3 & \hspace{-0.2cm}W4    &  Brightest X-ray source & $L_{0.5-7}^b$  & Brightest radio source & $F_{5.5}$ & $F_{9.0}$  & SFR$^c$ \\
  & (mag) & \hspace{-0.2cm}(mag) & \hspace{-0.2cm}(mag) &  \hspace{-0.2cm}(mag) &   (J2000) & $\left({\rm erg~s}^{-1}\right)$ &  (J2000) & ($\mu$Jy) & ($\mu$Jy) & $\left(M_{\odot}~{\rm yr}^{-1}\right)$    \\ %
    \hline\\[-7pt]
R-1 & 14.33 & 13.95 & 8.73 & 5.23 &  -- & $<10^{37}$ & 22:57:16.19, $-$41:05:17.7 & 267$\pm$15 & 145$\pm$8 & 0.04$\pm$0.01   \\
X-2 & 15.06 & 14.85 & 9.78 & 6.75 &  22:57:24.71, $-$41:03:44.1 & $6 \times 10^{39}$ & 22:57:24.70, $-$41:03:45.2 & 135$\pm$12 & 70$\pm$11 & 0.016$\pm$0.005   \\
C & 14.74 & 14.22 & 9.27 & 6.39 &  22:57:14.14, $-$41:02:49.2 & $6 \times 10^{37}$ & 22:57:12.87, $-$41:02:46.9 & 78$\pm$8 & 64$\pm$8 & 0.022$\pm$0.006   \\
    
\hline 
\vspace{-0.5cm}
\end{tabular}
\end{center}
\begin{flushleft} 
$^a$: see Figure 15 for the identification of the three off-nuclear star-forming regions;\\
$^b$: peak unabsorbed 0.5--7 keV luminosity during the two longer {\it Chandra} observations of 2002;\\
$^c$: average of the SFRs derived from W3 and W4, using the relations of \cite{cluver17}.\\

\end{flushleft}
\end{table*}

\section{A multiband look at star formation in NGC\,7424}
For a better understanding of the different star-forming properties of the environment around the X-1 and X-2 ULXs, we used archival data from three different bands: i) the {\it Wide-field Infrared Survey Explorer} ({\it WISE}; \citealt{wright10}) for the mid-IR; ii) a continuum-subtracted H$\alpha$ image from the 1.5-m telescope at the Cerro Tololo Inter-American Observatory (CTIO); iii) {\it GALEX} near-UV and far-UV images. 

{\it WISE} has four channels: W1 (3.4 $\mu$m), W2 (4.6 $\mu$m), W3 (12 $\mu$m) and W4 (22 $\mu$m). Three-color images of \{W1,W3,W4\} show three outstanding red clumps along the spiral arms (Figure 15). The data suggest that star formation is mostly concentrated in those three clumps. (Recall that in mid-IR images, redder colours indicate dusty star-forming regions, while bluer colours map the old stellar population). One of the three clumps corresponds to the star-forming complex around X-2 that we have analyzed in details in this work. Another clump corresponds to the brightest star cluster in NGC\,7424 \citep{larsen02} and the brightest radio source (R-1) identified by \cite{soria06}. The third clump, symmetrically located about 2$^{\prime}$ north-west of the nucleus, is resolved into a group of several young clusters and OB associations at the end of a spiral arm; it includes a fairly bright radio source, and a sub-Eddington X-ray binary, but no ULXs.  For short-hand notation, we have arbitrarily labelled the clumps ``X-2'', ``R-1'' and ``C'' in the three panels of Figure 15. The main properties of the three clumps are summarized in Table 6. Disk fragmentation into star-forming clumps is typically observed in high-redshift spirals \citep[{\it e.g.},][]{elmegreen07,dekel09,cava18} but is also sometimes seen in nearby galactic disks \citep[{\it e.g.},][]{fisher17,inoue18,larson20,dickinson22}.

The luminosity in the {\it WISE} W3 and W4 bands is a proxy for the total IR luminosity, and, hence, for the SFR. Using the observed brightness of the three clumps from the {\it AllWISE} catalog \citep{cutri14}, and the scaling of \cite{cluver17}, we find typical SFRs of a few $10^{-2} M_{\odot}$ yr$^{-1}$ for each of the three clumps (Table 6). The most active one is R-1, as also suggested by the optical brightness of the associated young star cluster \citep{larsen02}. All three clumps have W2$-$W3 colours $\approx$5 mag, which is typical of starburst galaxies (or starburst regions inside galaxies) and luminous infrared galaxies ({\it e.g.}, the classic diagram of \citealt{wright10}). None of the three clumps stands out in the H$\alpha$ and (especially) UV images, for the same reason why they do stand out in the {\it WISE} images: because dust removes photons from optical/UV bands and re-emits them in the mid-IR and far-IR.  A more detailed analysis of the star-formation properties of NGC\,7424 will be presented in a follow-up work currently in preparation, together with a comprehensive study of the radio sources (SNRs and H {\footnotesize II} regions) detected in our ATCA observations. Here, we only use the multiband data as an independent check for some of the luminosity estimates presented in Section 3.5.

The H$\alpha$ image\footnote{Downloaded from NED.} was taken from the 1.5-m CTIO telescope on 2000 September 16, with an exposure time of 1800 s; the 6568/28 filter was used for the narrow-band image, and an R-band image was used for continuum subtraction. The filter has a pivot wavelength $\lambda = 6575.51$ \AA, a Gaussian-equivalent FWHM of 36.1 \AA, and a rectangular equivalent bandpass of 31.2 \AA. Given the narrow width of the filter and the flux ratios measured in our VLT spectra, we estimate that the contamination of the [N {\footnotesize {II}}] lines is only around 5\% of the total flux in the band. We extracted the apparent flux from the region around the X-2 ULX, not including the nebular-only region to the north. We estimate an observed flux $F({\mathrm{H}}\alpha)_{\mathrm{tot,abs}} = (4.0\pm0.5) \times 10^{-14}$ erg cm$^{-2}$ s$^{-1}$, which compares well with our previous estimate of $F({\mathrm{H}}\alpha)_{\mathrm{tot,abs}} = (3.7\pm0.5) \times 10^{-14}$ erg cm$^{-2}$ s$^{-1}$ based on the long-slit VLT spectra. 

Next, we converted the {\it WISE} W4 magnitude into a flux density for the X-2 clump, following the prescriptions of \cite{wright10}\footnote{See also \url{https://wise2.ipac.caltech.edu/docs/release/allsky/expsup/sec4\_4h.html}}. We obtained a flux density $F_{\nu,22\mu{\rm m}} \approx 14.7 \mu$Jy, a corresponding monochromatic flux  $F_{22\mu{\rm m}} \approx 2.1 \times 10^{-12}$ erg cm$^{-2}$ s$^{-1}$, and a luminosity $L_{22\mu{\rm m}} \approx 2.9 \times 10^{40}$ erg s$^{-1}$. We then used the recipe of \cite{kennicutt09} to derive extinction-corrected H$\alpha$ luminosities: $L({\mathrm{H}}\alpha)_{\mathrm{unabs}} \approx L({\mathrm{H}}\alpha)_{\mathrm{abs}} + 0.02 L_{24\mu{\rm m}}$. We approximated the {\it Spitzer}/MIPS 24-$\mu$m luminosity (used to calibrate the Kennicutt relation) with the {\it WISE} 22-$\mu$m luminosity, which introduces only errors of a few per cent, negligible for the purpose of this simple exercise. 
The result is a $L({\mathrm{H}}\alpha)_{\mathrm{unabs}} \approx (1.1\pm0.1) \times 10^{39}$ erg s$^{-1}$ $\approx 2.2 L({\mathrm{H}}\alpha)_{\mathrm{abs}}$. This is in good agreement with the independent estimate of Section 3.5 based on the Balmer decrement. It converts to a reddening $E(B-V) \approx 0.33$ mag.


\begin{figure}
\centering
\includegraphics[width=\columnwidth]{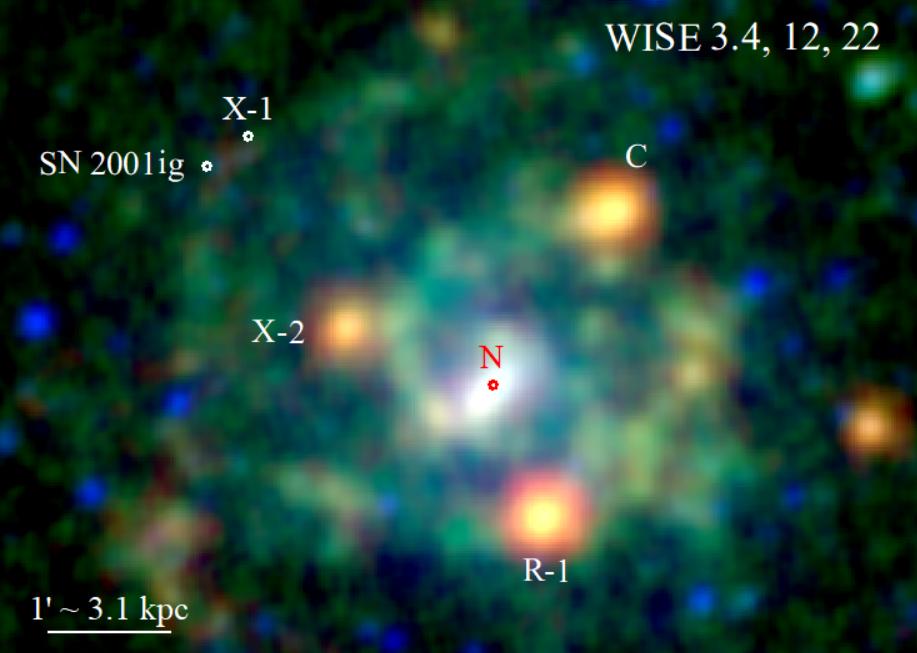}\\[10pt]
\includegraphics[width=\columnwidth]{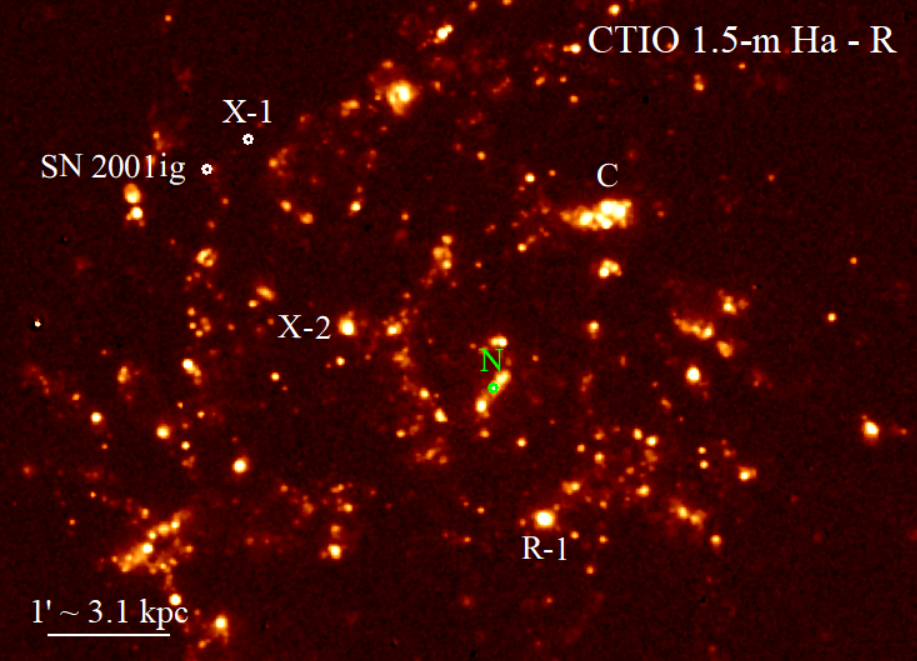}\\[10pt]
\includegraphics[width=\columnwidth]{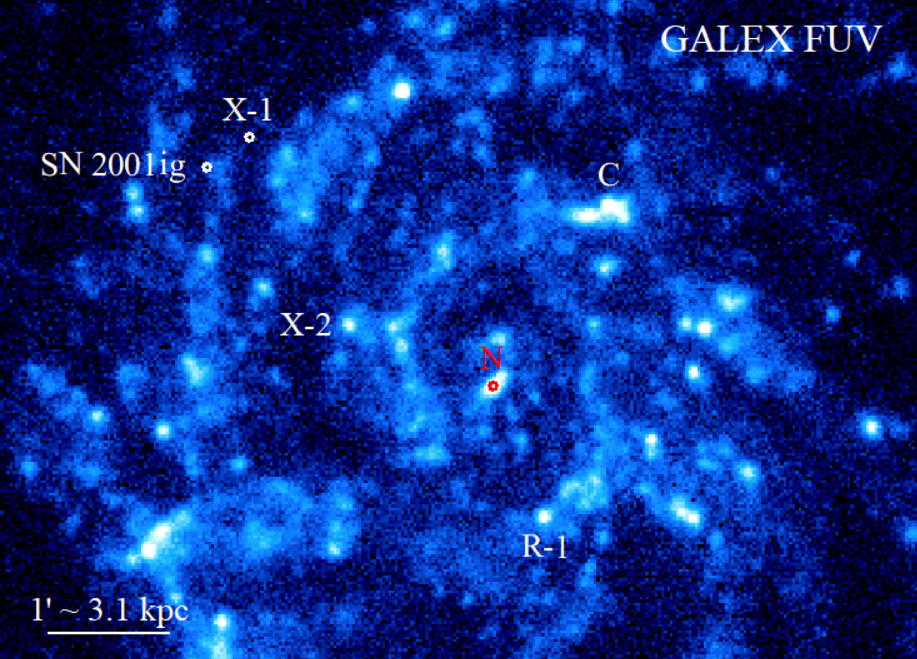}\\
\caption{Top panel: mid-IR three-colour image: red = {\it WISE} W4 band (22 $\mu$m), green = W3 (12 $\mu$m), blue = W1 (3.4 $\mu$m). The three dusty star-forming clumps in the spiral arms have been arbitrarily labeled for convenience (see also Table 6). For comparison, we also marked the position of the galactic nucleus, of SN\,2001ig, and of the X-1 ULX. The bluer colour of the nuclear region and the bar indicates an older population and weaker star formation in that region. Middle panel: continuum-subtracted H$\alpha$ image from the 1.5-m CTIO telescope. Bottom panel: GALEX far-UV image.
}
\label{wise}
\end{figure}


\section{Summary and Conclusions}

We took a second look at the two persistent ULXs in the face-on spiral NGC\,7424, two decades after their discovery. We showed that X-1 is located in a low density region, along a spiral arm but away from currently star-forming regions. We identified a blue, point-like optical counterpart. If this optical source is the donor star (rather than the irradiated accretion disk), it is consistent with a blue supergiant with an age $\approx$28 Myr. In this case, this ULX belongs to the same ULX class as, for example, NGC\,7793 P13 \citep{motch14}. 

X-1 has been found in a bright state in all X-ray observations (three {\it Chandra}, two {\it Swift} and one {\it SRG}/eROSITA observations) over a span of 20 years, at $L_{\rm X} \approx 5$--$9 \times 10^{39}$ erg s$^{-1}$, with no obvious spectral state transitions.
By contrast, X-2 has more unusual properties. It is located in one of the three most active star-forming clumps in the whole galactic disk, a region with a size of $\approx$100 pc $\times$ 150 pc, populated by young star clusters. Our VLT spectroscopic study shows strong line emission from a UV-photoionized H {\footnotesize {II}} region, as expected, but also strong nebular He {\footnotesize {II}} $\lambda$4686 emission from the location around the X-ray source---not from the brightest and youngest part of the star-forming clump, which is located $\approx$50 pc south of the ULX.

Nebular He {\footnotesize {II}} $\lambda$4686 emission ({\it i.e.}, not from Wolf-Rayet winds) is a hallmark of photo-ionization by soft X-ray photons, and has been found around several ULXs. For this He$^{++}$ region, we estimated a luminosity $L(4686) \approx 8 \times 10^{36}$ erg s$^{-1}$. This makes the NGC\,7424 X-2 nebula about 3 times more luminous than the well-known He$^{++}$ nebula around Holmberg II X-1. For gas at a temperature of $\approx10^4$ K, ionized by soft X-ray photons, an approximate conversion between the long-term-average (over the recombination timescale of a few 1000 yr) ionizing photon rate $Q({\rm He}^+)$ (in the 54--300 eV band) and the He {\footnotesize {II}} $\lambda$4686 luminosity is $L(4686) \approx 1.02 \times 10^{-12} Q({\rm He}^+)$ erg s$^{-1}$ ({\sc cloudy} code: \citealt{ferland98}; see also \citealt{osterbrock06,kaaret04,pakull89,pakull86}). For X-2 in the high state (2002 June and 2020 December), different X-ray fitting models predict a range of ionizing photon fluxes, between $\approx$4--20 $\times 10^{48}$ s$^{-1}$, corresponding to $L(4686) \sim 4$--20 $\times 10^{36}$ erg s$^{-1}$. The luminosity inferred from the VLT spectra (with various assumptions on slit losses, spatial extent of the nebula, dust extinction) is $L(4686) \approx 8 \times 10^{36}$ erg s$^{-1}$, in the middle of the predicted range.

A second point of interest of NGC\,7424 X-2 is that the system has a younger age than most other ULXs. From the Balmer emission lines, we estimate an age of $(4.4 \pm 0.3)$ Myr for the stellar population around the ULX. The stellar progenitor of the ULX must have collapsed after less than this age. This implies a progenitor mass $M \gtrsim 47 M_{\odot}$ (from the Padova isochrones at solar metallicity, \citealt{bressan12}). It does not, however, imply that the compact object is a black hole. The example of a neutron star found in the young Galactic cluster Westerlund 1 (age $\lesssim$5 Myr) suggests that binary evolution can lead to a supernova and neutron star formation even from progenitors above 40 $M_{\odot}$ \citep{schneider21,belczynski08}.
With such a young age, the donor star of X-2 cannot be a blue supergiant (unlike the X-1 donor), and is instead either a main-sequence O star or a Wolf-Rayet.

A third interesting feature of X-2 is that it was seen once in a lower-luminosity state ($L_{\rm X} \approx 6 \times 10^{38}$ erg s$^{-1}$), with a hard power-law-like spectrum in {\it Chandra}'s 0.5--7 keV band. A simple assumption that this property corresponds to the bright end of the hard state of an accreting black hole ($L_{\rm X} \lesssim 0.1 L_{\rm Edd}$) would imply a black hole mass $\gtrsim$50 $M_{\odot}$. On the other hand, super-critical accretion onto strongly magnetized (young) neutron stars also produces a hard spectrum, from fan-beam emission in the accretion column. The softening of the source with brightness in the subsequent two {\it Chandra} observations may be explained by a higher contribution from the inner disk and a denser disk outflow along our line of sight, as the accretion rate increased \citep{gurpide21,gurpide21a}. 

A fourth interesting property of X-2, or at least of the environment around X-2, is the presence of a strong (2.5 times the luminosity of Cas A at 5.5 GHz), unresolved radio source at the southern end of the star-forming clump. Based on the observed Balmer emission, we argued that free-free emission from gas in the main star cluster is expected to contribute a non-negligible component, but cannot be the dominant process. The steep spectrum suggests a dominant optically-thin synchrotron component. It could be a young (age $\lesssim$1000 yr) SNR, completely unrelated to the ULX; however, the massive star cluster is too young (age $<$3 Myr) to have a substantial SN rate. Or it could be the radio hot spot/radio lobe of a jet powered by the ULX. This would be analogous to the jets in NGC\,7793 S26 \citep{soria10} and in NGC\,6946 X-1 (T.~Beuchert et al., in prep.), where the peak radio synchrotron emission is displaced from the compact object. The non-detection of a symmetrically placed radio lobe north of NGC\,7424 X-2 could be explained by the much lower ISM density in that region. 

If the non-thermal radio source is caused by shocks from a ULX jet, we expect the presence of broad optical emission lines from shocked-ionized gas. The H$\beta$ luminosity expected from the shocked gas, for typical ULX bubble parameters, is $L({\mathrm{H}}\beta) \approx 2.2$--$2.5 \times 10^{-3} P_{\rm jet}$ \citep{allen08}. If we assume $P_{\rm jet} \approx L_{\rm X} \approx 7 \times 10^{39}$ erg s$^{-1}$, the shock-ionized component of the Balmer emission is only $\approx$10\% of the Balmer luminosity from the star cluster (Section 3.5). This weaker component is consistent with the broader wings marginally detected in the strongest lines (Section 3.4). An objection to the ULX jet scenario is that two strong radio sources are also associated with the other two main starburst clumps in NGC\,7424, although they do not contain currently active ULXs. In summary, we do not have enough information to rule out any of those scenarios for the origin of this intriguing radio source near X-2. We are planning deeper ATCA observations in 2024 to put a stronger constraint to the spatial extent and spectral index, and to look for possible northern radio lobe.

\section*{Acknowledgements}
RS acknowledges grant number 12073029 from the National Natural Science Foundation of China (NSFC). 
RS also acknowledges support and hospitality from the Observatoire de Strasbourg during part of this work. 
TDR is supported by a IAF-INAF Research Fellowship.
This research benefitted from discussions at the International Space Science Institute in Bern, through the team led by L.~Oskinova: Multiwavelength View on Massive Stars in the Era of Multimessanger Astronomy.
We thank the anonymous referee for their suggestions, which have improved the original manuscript. 
We thank Stuart Ryder, who provided us with the calibrated set of stacked, flat-fielded Gemini images from his program GS-2004B-Q-6.
We thank Tobias Beuchert, Hua Feng, Andr\'es Gurpide, Matt Middleton, Lida Oskinova, Ciro Pinto, Sabela Reyero Serantes, Beverly Smith, Alexandr Vinokurov and Dom Walton for discussions about SNRs and ionized bubbles, and Alister W.~Graham for discussions on the use of {\it WISE} data and of spiral galaxy properties. 

This research has made use of data obtained from the {\it Chandra} Data Archive and the {\it Chandra} Source Catalog, and software provided by the Chandra X-ray Center (CXC) in the {\sc ciao} application package. Furthermore, we used data from eROSITA, the soft X-ray instrument aboard {\it SRG}, a joint Russian-German science mission supported by the Russian Space Agency (Roskosmos), in the interests of the Russian Academy of Sciences represented by its Space Research Institute (IKI), and the Deutsches Zentrum f\"ur Luft- und Raumfahrt (DLR). The {\it SRG} spacecraft was built by Lavochkin Association (NPOL) and its subcontractors, and is operated by NPOL with support from the Max Planck Institute for Extraterrestrial Physics (MPE). The development and construction of the eROSITA X-ray instrument was led by MPE, with contributions from the Dr. Karl Remeis Observatory Bamberg \& ECAP (FAU Erlangen-Nuernberg), the University of Hamburg Observatory, the Leibniz Institute for Astrophysics Potsdam (AIP), and the Institute for Astronomy and Astrophysics of the University of T\"ubingen, with the support of DLR and the Max Planck Society. The Argelander Institute for Astronomy of the University of Bonn and the Ludwig Maximilians Universität Munich also participated in the science preparation for eROSITA. Moreover, we used observations made with the NASA/ESA {\it Hubble Space Telescope}, and obtained from the Hubble Legacy Archive, which is a collaboration between the Space Telescope Science Institute (STScI/NASA), the Space Telescope European Coordinating Facility (ST-ECF/ESA) and the Canadian Astronomy Data Centre (CADC/NRC/CSA). This work is also based on observations obtained at the international Gemini Observatory, a program of NSF’s NOIRLab, which is managed by the Association of Universities for Research in Astronomy (AURA) under a cooperative agreement with the National Science Foundation on behalf of the Gemini Observatory partnership: the National Science Foundation (United States), National Research Council (Canada), Agencia Nacional de Investigaci\'{o}n y Desarrollo (Chile), Ministerio de Ciencia, Tecnolog\'{i}a e Innovaci\'{o}n (Argentina), Minist\'{e}rio da Ci\^{e}ncia, Tecnologia, Inova\c{c}\~{o}es e Comunica\c{c}\~{o}es (Brazil), and Korea Astronomy and Space Science Institute (Republic of Korea). 
The Australia Telescope Compact Array is part of the Australia Telescope National Facility ({\url {https://ror.org/05qajvd42}}) which is funded by the Australian Government for operation as a National Facility managed by CSIRO. We acknowledge the Gomeroi people as the Traditional Owners of the ATCA observatory site, the Taurini as the Traditional Owners of the INAF-OATo site, and the Triboces as the Traditional Owners of the Observatoire de Strasbourg (Argentorate) site.
Moreover, we used data products from the {\it Wide-field Infrared Survey Explorer}, which is a joint project of the University of California, Los Angeles, and the Jet Propulsion Laboratory/California Institute of Technology, funded by the National Aeronautics and Space Administration. 
We used {\sc iraf} software for part of the optical analysis: {\sc iraf} is distributed by the National Optical Astronomy Observatory, which is operated by the Association of Universities for Research in Astronomy (AURA) under a cooperative agreement with the National Science Foundation.

\section*{Data Availability}
The {\it Chandra}, {\it Swift}, {\it HST}, Gemini, {\it WISE}, VLT and ATCA datasets used for this work are all available for download from their respective public archives. 

\bibliographystyle{mnras}
\bibliography{main}

\bsp	
\label{lastpage}

\end{document}